\documentclass[a4paper,fleqn,usenatbib]{mnras}

\usepackage[T1]{fontenc}
\usepackage{ae,aecompl}

\usepackage{graphicx}
\usepackage{amsfonts}
\usepackage{natbib}
\usepackage{amssymb, amsmath, amsbsy}
\usepackage[dvips]{color}

\defcitealias{bs16}{BS16}
\defcitealias{sb15}{SB15}
\defcitealias{cm77}{CM77}

\newcommand{\kms}{\,km\,s$^{-1}$} 
\newcommand{\kmss}{\,\mathrm{km\,s}^{-1}}
\newcommand{\msolar}{M_\odot} 
\newcommand{\kelvin}{\,\mathrm{K}}

\newcommand{\kB}{k_\mathrm{B}}
\newcommand{\mh}{m_\mathrm{H}}

\newcommand{\fig}{Figure \ref}
\newcommand{\eqn}{Equation \ref}
\newcommand{\tab}{Table \ref}
\newcommand{\sect}{Section \ref}
\newcommand{\bq}{\begin{equation}}
\newcommand{\eq}{\end{equation}}

\newcommand{\mach}{\mathcal{M}}

\newcommand{\qs}{\hat{q}_\mathrm{s}}
\newcommand{\fS}{f_\mathrm{s}}

\newcommand{\vrel}{v_\mathrm{rel}}

\newcommand{\mcloud}{M_\mathrm{c}}
\newcommand{\rcloud}{R_\mathrm{c}}
\newcommand{\lcloud}{L_\mathrm{c}}

\newcommand{\Ncloud}{N_\mathrm{c}}
\newcommand{\pram}{P_\mathrm{ram}}

\newcommand{\etas}{\eta_\mathrm{s}}
\newcommand{\taus}{\tau_\mathrm{s}}
\newcommand{\betas}{\beta_\mathrm{s}}

\newcommand{\rhocloud}{\rho_\mathrm{c}}
\newcommand{\Tcloud}{T_\mathrm{c}}
\newcommand{\pcloud}{P_\mathrm{c}}

\newcommand{\lkh}{\lambda_\mathrm{KH}}
\newcommand{\fkh}{f_\mathrm{KH}}

\newcommand{\mvir}{M_\mathrm{vir}}
\newcommand{\rvir}{R_\mathrm{vir}}
\newcommand{\Tvir}{T_\mathrm{vir}}
\newcommand{\logmvir}{\log(M_\mathrm{vir}/M_\odot)}
\newcommand{\machre}{\mathcal{M}_\mathrm{re}}

\newcommand{\vcirc}{v_\mathrm{cir}}
\newcommand{\siggal}{\sigma_\mathrm{gal}}
\newcommand{\vwind}{v_\mathrm{w}}
\newcommand{\nngb}{N_\mathrm{ngb}}

\newcommand{\rhoa}{\rho_\mathrm{a}}
\newcommand{\Ta}{T_\mathrm{a}}
\newcommand{\Za}{Z_\mathrm{a}}
\newcommand{\pa}{P_\mathrm{a}}
\newcommand{\namb}{n_\mathrm{a}}

\newcommand{\mlr}{\dot{M}_\mathrm{c}}

\newcommand{\delm}{\Delta M}
\newcommand{\delp}{\Delta p}

\newcommand{\delE}{\Delta E}
\newcommand{\delv}{\Delta v}

\newcommand{\delt}{\Delta t}

\newcommand{\flim}{f_\mathrm{lim}}
\newcommand{\ft}{f_\mathrm{t}}
\newcommand{\fL}{f_\mathrm{L}}

\newcommand{\figpath}{./figures}

\title[Implementation of PhEW]{A New Model For Including Galactic
Winds in Simulations of Galaxy Formation II: Implementation of PhEW in
Cosmological Simulations}

\author[S. Huang et al.]{
Shuiyao Huang$^{1}$\thanks{E-mail:shuiyao@umass.edu},
Neal Katz$^{1}$, J'Neil Cottle$^{2}$, Evan Scannapieco$^{2}$, \newauthor
Romeel Dav\'e$^{3,4,5}$, \& David H. Weinberg$^{6,7}$
\\$^1$ Astronomy Department, University of Massachusetts, Amherst, MA
01003,
USA
\\$^2$ School of Earth and Space Exploration, Arizona State
University, P.O. Box 871404, AZ 85287-1404, USA
\\$^3$ Institute for Astronomy, Royal Observatory, University of
Edinburgh, Edinburgh EH9, 3HJ, UK
\\$^4$ University of the Western Cape, Bellville, Cape Town 7535,
South Africa 
\\$^5$ South African Astronomical Observatories, Observatory, Cape
Town 7925, South Africa
\\$^6$ Astronomy Department and CCAPP, Ohio State University,
Columbus, OH 43210, USA
\\$^7$ Institute for Advanced Study, Princeton, NJ 08540, USA
}

\date{Accepted 0000 October 00. Received 0000 October 00; in original
form 0000 October 00}

\pubyear{2021}

\begin{document}
\label{firstpage}
\pagerange{\pageref{firstpage}--\pageref{lastpage}} \pubyear{0000}
\maketitle


\begin{abstract}

Although galactic winds play a critical role in regulating galaxy formation,
hydrodynamic cosmological simulations do not resolve the scales that govern the
interaction between winds and the ambient circumgalactic medium (CGM). We
implement the Physically Evolved Wind (PhEW) model of Huang et al. (2020) in
the \textsc{gizmo} hydrodynamics code and perform test cosmological simulations with
different choices of model parameters and numerical resolution. PhEW adopts an
explicit subgrid model that treats each wind particle as a collection of clouds
that exchange mass and metals with their surroundings and evaporate
by conduction and hydrodynamic instabilities as calibrated on much higher 
resolution cloud
scale simulations. In contrast to a conventional wind algorithm, we find that PhEW results are robust to numerical resolution and implementation details
because the small scale interactions are defined by the model itself.
Compared to our previous wind simulations with the same resolution, our PhEW
simulations are in
better agreement with low-redshift galactic stellar mass functions at $M_* < 10^{11}M_\odot$
because PhEW particles shed mass to the CGM before escaping low mass halos.
PhEW radically alters the CGM metal distribution because PhEW particles
disperse metals to the ambient medium as their clouds dissipate, producing a
CGM metallicity distribution that is skewed but unimodal and is similar between
cold and hot gas. While the temperature distributions and radial profiles of
gaseous halos are similar in simulations with PhEW and conventional winds,
these changes in metal distribution will affect their predicted UV/X-ray
properties in absorption and emission.

\end{abstract}

\section{Introduction}
\label{sec:introduction}
Galactic winds have been ubiquitously observed in galaxies at both low
and high redshifts, and they are critical to galaxy formation and evolution.
Simulations calibrated to match these observations predict that a large amount of galactic material is ejected as a wind before reaccreting to either form stars or be ejected once again \citep{oppenheimer10, angles-alcazar17}.
Current cosmological hydrodynamic simulations of galaxy formation
employ a variety of sub-grid models \citep[e.g.][]{springel03,
oppenheimer06, stinson06, dallavecchia08, agertz13, schaye15, mufasa,
tremmel17, pillepich18a, simba, huang20a} that artificially launch
galactic winds, but the results are sensitive to numerical
resolution and the exact sub-grid model employed \citep{huang19,huang20a}.
Simulations without these sub-grid wind models \citep[e.g.][]{fire2, kim15,
martizzi16}
allow winds to occur ``naturally'', but these simulations may not resolve 
the scales necessary to resolve the important known physical processes
\citep{sb15, bs16,schneider17,mccourt18, phewi}.
Hence, modelling galactic winds accurately remains a theoretical
challenge for even the most refined high-resolution simulations of
galaxies (see \citet{naab17}, for a review).
Even if one were able to accurately model the formation of galactic winds,
the subsequent propagation in
galactic haloes depends on a complicated interplay of many physical
processes that occur on a wide range of physical scales that cannot
be simultaneously resolved in a single simulation. For example, to
robustly model the propagation and disintegration of moving clouds in
various situations requires cloud-crushing simulations with at least
sub-parsec scale resolution \citep{schneider17, mccourt18}, which is
orders of magnitudes below the resolution limits of cosmological
simulations.
Furthermore, most cosmological hydrodynamic simulations concentrate their resolution in the dense, star-forming regions of galaxies and thus have lower resolution in the circumgalactic medium \citep[CGM, but see ][]{hummels19, mandelker19, peeples19, suresh19, vandevoort19}.
To date, cosmological simulations do not include physically motivated sub-grid models for galactic wind evolution, which are required to capture these small-scale physical processes.

Owing to the complicated nature of galactic winds, the sub-grid wind
models used to launch the winds vary widely among different cosmological 
simulations. The
results from these simulations are sensitive to the details of the
wind implementation and the wind model parameters \citep{huang20a}.
They are resolution-dependent due to the poorly understood interaction
between the numerical resolution and the numerical method used to solve the
hydrodynamics. For example, in conventional wind algorithms, one needs to solve for the
evolution of a single wind particle moving supersonically in the neighborhood of gas particles whose properties may be quite different from those of the wind particle. The interactions between the wind particle and the ambient medium, such as shocks, are unresolved in the simulation. Details of the momentum exchange between the wind particle and the neighboring particles and the entropy generated in the process completely rely on the numerical methods, e.g., artificial viscosity, that are different among different simulations and are almost certainly not physically accurate.
\citet{huang20a} found that a small change in the initial wind velocities may result
in large differences in the amounts of wind re-accretion.
Furthermore, the results are not only sensitive to how the winds are
launched from galaxies but also to their subsequent evolution and
interaction with the surrounding medium.

Inadequate resolution and over-simplified wind models may also affect
wind material in the CGM
and the intergalactic medium (IGM). Observational evidence shows that
the CGM is composed of multi-phase gas on a range of physical scales
\citep{tumlinson17}, many of which are not resolvable in cosmological
simulations \citep{mccourt18}. Current wind models do not
self-consistently account for the multi-phase sub-structure of the
wind particles in simulations.

In \citet{phewi}, we developed an analytic model, Physically Evolved
Winds (PhEW), which predicts the evolution of a cold cloud that travels
at supersonic speeds through an ambient medium.
The PhEW model explicitly calculates various properties of the
evolving cloud, such as the cloud mass, $\mcloud$, the cloud velocity
relative to the ambient medium, $\vrel$, the cloud density,
$\rhocloud$, and the cloud temperature $\Tcloud$ as functions of time during the cloud's evolution.
The model includes physical processes such as shocks, hydrodynamic
instabilities and thermal conduction simultaneously in the calculation,
and we tuned the cloud evolution model to match the numerical
results from high resolution cloud-crushing simulations of
\citet{bs16} (hereafter, \citetalias{bs16}) with thermal conduction
and radiative cooling under many conditions.

Cloud-crushing simulations agree well when they include the same physics and adopt a consistent definition of cloud mass. For example, \citet{schneider17} carried out simulations of the pure hydrodynamic case with radiative cooling, using the Cholla code. This is a GPU-enabled code that allows for the use of a large static mesh, as opposed to the smaller, frame-changing domain simulated by the \textsc{flash} code in \citet{bs16}. Nevertheless, as discussed in detail in \citet{schneider17}, the differences in mass loss and acceleration between the two studies are relatively minor and related to details of the cooling adopted. 

Similarly, \citet{li19} explored a wide range of model parameters with the \textsc{gizmo} code, finding results that agree with the \textsc{flash} results on which PhEW is based when they adopt the same choice of unknown physics such as the conduction coefficient, magnetic field structure, etc. An important positive aspect of PhEW is that it can allow for full cosmological simulations that incorporate different models of such small-scale physics, such as the importance of hydrodynamic instabilities and thermal conduction as encapsulated in the corresponding parameters that control their strength.

In this paper, we describe a method to implement our analytic PhEW
model into cosmological simulations. Though in this paper we detail
implementation in the \textsc{gizmo}
\citep{gizmo} hydrodynamics code, the method could be generalised to any
cosmological hydrodynamic simulation including galactic winds. Even if the 
hydrodynamic method does not explicitly involve the use of gas particles,
wind particles could be temporarily created, propagated using the PhEW model,
and then destroyed in a way that conserved all the important quantities. 
In our approach, we eject wind particles as in
our previous particle-based or mass-conserving simulations \citep{oppenheimer06, huang20a} but follow their
evolution in the ambient
CGM differently. We model each wind particle
as an ensemble of identical cloudlets that travel
at the same speed, with each of the individual cloudlets evolved using the
PhEW model. As a result, PhEW allows wind particles to exchange mass and metals with its neighboring particles based on a physically motivated prescription, therefore enabling metal mixing between the wind and the ambient medium. This metal mixing is not implemented in our previous simulations.

In contrast to our previous simulations of the CGM \citep{oppenheimer09, dave10, oppenheimer12, ford13, ford14, ford16}, PhEW incorporates an approximate but explicit description of the physics on scales that cannot be resolved by hydrodynamic simulations, yields predictions that are less sensitive to the hydrodynamic resolution (as we show below), and allows a physically realistic treatment of metal mixing between winds and the ambient CGM.

Recent simulations indicate that under certain physical conditions the cold gas mass in outflowing clouds could even increase, as opposed to decrease, with time \citep[e.g.][]{marinacci10, armillotta16, gronke18, li19}. For example, \citet{marinacci10} studied the infall of cold clouds onto the Milky Way and showed that the mixed material trailing relatively slow-moving (75\kms) clouds moving through the $2\times10^6$ K disk-CGM interface, displayed a significant enhancement in cooling.  This is because, in this case, the mixed gas is near the $\sim 10^5$ K peak of the cooling curve, which can lead to significant amounts of condensation in the cloud's wake.  Similarly, \citet{gronke18} showed that transonic ($\mach=1.5$) clouds were able to induce significant cooling in the exterior medium, if the cooling time of the cloud was sufficiently small. This increase in mixed cooling gas, labeled as cloud “growth” by \citet{gronke18}, is captured in our simulations by regular gas particles that accrete material from the PhEW particles. We can approximate this process in this way because such mixed gas travels much more slowly than the gas in the core of the cloud tracked by our sub-grid model. 

The paper is organised as follows.
In \sect{sec:method}, we provide details of the numerical
implementation of the PhEW model in cosmological simulations and
discuss the assumptions and choices we make.
In \sect{sec:simulation}, we describe several test simulations that we
perform to study the effects of the PhEW model.
In \sect{sec:phewparts}, we study the properties and behaviours of
individual PhEW particles in our test simulations and how they depend
on the wind parameters and numerical resolution.
In \sect{sec:galaxies}, we analyse the stellar and gas properties of
galaxies in these simulations. We demonstrate how galaxies acquire
their mass and metals differently with and without the PhEW model.
In \sect{sec:discussion}, we summarise our main results.

\section{Implementation}
\label{sec:method}

In this section, we describe the implementation of the PhEW model into
hydrodynamic simulations of galaxy formation as a sub-grid recipe for
evolving galactic winds. We will describe the method based on
our own simulations that use a particle-based hydro solver, e.g.,
smoothed particle hydrodynamics \citep[SPH,][]{oppenheimer06, simba, huang20a} or \textsc{gizmo} with meshless finite-mass
\citep[MFM,][]{gizmo}. 

Here, we focus on the propagation of wind particles in galactic haloes
after they have escaped the galaxies from which they were ejected.
\citet{huang20a} describe the ejection of wind particles in our
cosmological simulations in detail.

To summarise, at each time-step we
find gas particles that are above a density threshold,
$\rho_\mathrm{SF}$, equivalent to a hydrogen number density of
$n_\mathrm{H}=0.13\,\mathrm{cm}^{-3}$, and treat them as star-forming particles.
In addition, we use an on the fly friends-of-friends algorithm 
to identify the dark matter haloes to which these star-forming particles
belong and calculate the velocity dispersion, $\siggal$, of those haloes.
Then we select star-forming particles randomly to eject as wind
particles at a rate proportional to the mass loading factor $\eta_\mathrm{w}$,
giving each of them an initial wind velocity $\vwind$. Both $\eta_\mathrm{w}$
and $\vwind$ are functions of $\siggal$ \citep{huang20a}:

\begin{equation}
\eta_\mathrm{w}=\left\{                                                                                                                  
\begin{split}                                                                                                                 
  &\alpha_\eta\left(\frac{150\kmss}{\siggal}\right)\left(\frac{\sigma_\mathrm{ezw}}{\siggal}\right)^{\beta_\eta} &(\siggal < \sigma_\mathrm{ezw})\\
  &\alpha_\eta\left(\frac{150\kmss}{\siggal}\right) &(\siggal \geqslant \sigma_\mathrm{ezw})
\end{split}                                                                                                                   
\right.                                                                                                                       
\end{equation} 
where we choose the free parameters, $\alpha_\eta=0.1$, $\beta_\eta=4.0$ and $\sigma_\mathrm{ezw}=106\kmss$, for all simulations in this paper. We also allow $\eta_\mathrm{w}$ to change with redshift as $\eta_\mathrm{w}\propto (1+z)^{1.3}$ at $z<4$ and use a constant factor of $5^{1.3}$ at $z>4$.

The formula for the initial wind velocity, $\vwind$, is
\begin{equation}
  \vwind = \alpha_v\siggal\sqrt{f_\mathrm{L}}\left(\frac{\siggal}{50\ \mathrm{km\ s^{-1}}}\right)^{\beta_v},
  \label{eqn:vwind}
\end{equation}
where $\fL$ is a metallicity dependent ratio between the galaxy luminosity and the
Eddington luminosity. We adopt the \citet{oppenheimer06} formula for $\fL$:
\begin{equation}
  f_\mathrm{L} = f_{\mathrm{L};\odot}\times 10^{-0.0029(\log Z_\mathrm{gal} + 9 )^{2.5} + 0.417694},
\end{equation}
where we choose $f_{\mathrm{L};\odot} = 2$ as in \citet{oppenheimer06} so that $f_\mathrm{L}$ typically ranges from 1.05 to 2.
We also use the mass weighted average metallicity $Z_\mathrm{gal}$ of the host galaxy to compute $\fL$.
In this paper, we use $\alpha_v=3.5$, $\beta_v=0.6$ for all simulations.

After ejecting the wind particles, we
temporarily let each of them move out freely, experiencing only gravitational accelerations, until the
ambient density decreases below $\rho_\mathrm{th} = 0.1\rho_\mathrm{SF}$.
At this time we start modelling the wind particle as a PhEW particle
using the analytic
method from \citet{phewi}.


We assume that each PhEW particle of mass $M_i$ and velocity $v_i$
represents $\Ncloud$ identical cloudlets, each with the same initial
mass $\mcloud$ and velocity $v_i$, which is the same as the wind particle 
velocity. The
number of cloudlets in a PhEW particle, $\Ncloud = M_i / \mcloud$,
depends on the mass resolution ($M_i$) and the choice of the model
parameter $\mcloud$. These cloudlets evolve identically and
independently according to the analytic model. Therefore, the PhEW
particle that represents these cloudlets has the same phase properties
as each individual cloudlet.

At each time-step, we evaluate the properties of the ambient medium
using the same kernel as that used for the hydrodynamics solver. 
In a SPH or a MFM simulation, one defines the kernel as a normalised,
spherically symmetric function $W(r/h_i), (r < h_i)$ for each gas particle,
where $r$ is the distance from the particle and the $h_i$ is the softening
length of the particle. We determine the softening length $h_i$ of
each particle by specifying a fixed number of neighbouring gas
particles, i.e., $\nngb=128$, within the kernel. We determine the ambient
density $\rhoa$, the ambient temperature $\Ta$, and the ambient
metallicity $\Za$ using the kernel smoothed values from the neighbouring particles:
\bq
\rhoa = \sum_j^{\nngb} m_jW_{ij},
\label{eqn:rho_ambient}
\eq
\bq
\Ta = \left(\frac{\mu \mh}{\kB}\right)\frac{\sum_j^{\nngb}
m_ju_jW_{ij}}
{\sum_j^{\nngb}m_jW_{ij}},
\label{eqn:temperature_ambient}
\eq
and
\bq
\Za = \frac{\sum_j^{\nngb} m_jZ_jW_{ij}}
{\sum_j^{\nngb}m_jW_{ij}},
\label{eqn:metallicity_ambient}
\eq
where $W_{ij} \equiv W(|\mathbfit{r}_i - \mathbfit{r}_j|/h_i)$, $m_j$,
$u_j$ and $Z_j$ are the mass, specific internal energy, and metallicity
of a neighbouring particle $j$, and $\mu$, $\mh$ and $\kB$ are the atomic
weight, hydrogen mass, and the Boltzmann constant, respectively. One
can derive the thermal pressure of the ambient medium from these
properties as
\bq
\pa = \frac{\rhoa\kB\Ta}{\mu\mh}.
\label{eqn:pressure_ambient}
\eq

We determine the relative velocity between the cloudlets and the
ambient medium similarly, but only include the neighbouring particles
that move against the PhEW particle as we are only concerned with
converging flows where shocks can develop:
\bq
\vrel = \left|\frac{\sum_j^{\nngb}
m_jV_{ij}|\mathbfit{v}_i-\mathbfit{v}_j|W_{ij}}
{\sum_j^{\nngb}m_jV_{ij}W_{ij}}\right|,
\label{eqn:vrel}
\eq
where
\bq
V_{ij} =
\begin{cases}
1, & \mathbfit{v}_i \cdot (\mathbfit{v}_j - \mathbfit{v}_i) \ge 0;\\
0, & \mathbfit{v}_i \cdot (\mathbfit{v}_j - \mathbfit{v}_i) < 0.
\end{cases}
\eq

When a PhEW particle uses its neighboring particles to determine the ambient properties, it excludes all other PhEW particles but does not explicitly exclude the star-forming gas. However, since we only allow a PhEW particle to interact with the ambient gas once it has traveled a distance from its host galaxy and recouple a PhEW particle when it hits a galaxy, it is rare for a PhEW particle to have any star-forming neighbors. For example, the hydrodynamic softening length, i.e., size of the three dimensional kernel, at $\rho_\mathrm{th}$ (where the PhEW particles are initialized) is around 15 kpc, which is smaller than or at least comparable to the distance from the galaxy where it is launched.

At the first time-step when a wind particle becomes a PhEW
particle, we initialise the cloudlet properties as if they have just been
swept by the cloud-crushing shock. We set the initial temperature of
the cloudlet to be $10^4\kelvin$ assuming the shock is isothermal, and
set the initial pressure $\pcloud$, density $\rhocloud$, radius
$\rcloud$ and length $\lcloud$ of each cloudlet as follows
\citep{phewi}.
\bq
\pcloud =
\pa \left [ \frac{2\gamma}{\gamma+1}\mach^2 -
\frac{\gamma-1}{\gamma+1} \right]
\etas\taus,
\label{eqn:pressure_cloud}
\eq
where $\mach$ is the Mach number of the ambient flow relative to the
cloudlet, $\etas$ and $\taus$ are the corrections to the jump
conditions for density and temperature owing to thermal conduction
with
\bq
\label{eqn:jc_density}
\etas = \frac{5(1+\betas) +
\sqrt{9+16\qs+5\betas(5\betas-6)}}{8(1-\qs+5\betas)}
\eq
and
\bq
\label{eqn:jc_temperature}
\taus = \frac{1}{2} - \frac{4\qs}{3} +
\frac{1}{6}(1+\betas)\sqrt{9+16\qs+5\betas(5\betas-6)} +
\frac{5}{6}\betas(\betas + 6),
\eq
where $\qs$ is a parameter chosen as 0.9 and $\betas$ is defined as
$\betas \equiv 1/(\gamma \mach^2)$.

The density, radius, and length of the cloudlet are
\bq
\rhocloud = \frac{\mu\mh}{10^4\kelvin \cdot \kB}\pcloud,
\label{eqn:density_cloud}
\eq
\bq
\rcloud = \left( \frac{\mcloud}{2\pi\rhocloud}\right)^{\frac{1}{3}},
\label{eqn:radius_cloud}
\eq
and
\bq
\lcloud = 2\rcloud.
\label{eqn:length_cloud}
\eq

At each succeeding time-step, we first obtain the properties of the
ambient medium using \eqn{eqn:rho_ambient} to \eqn{eqn:vrel}, and then
calculate how each cloudlet will evolve according to the analytic
model. Namely, we find the mass loss rate $\mlr$, the deceleration
$\dot{\mathbfit{v}}_i$, and the changes in the geometric parameters of
the cloudlet, i. e., $\dot{\rcloud}$ and $\dot{\lcloud}$, as described in
\citet{phewi}.

We assume that the cloudlets lose their mass to the ambient medium
owing to hydrodynamic instabilities and also thermal evaporation. We
calculate the total mass loss rate of a cloudlet combining these two
effects as (Equation 43 from \citet{phewi})
\bq
\label{eqn:mlr}
\dot{M}_\mathrm{c} = \dot{M}_\mathrm{c,KH}\exp(-\rcloud/\lkh) +
\dot{M}_\mathrm{c,ev},
\eq
where $\dot{M}_\mathrm{c,KH}$ and $\dot{M}_\mathrm{c,ev}$ are mass
loss rate from hydrodynamic instabilities and evaporation alone,
respectively.
\bq
\dot{M}_\mathrm{c;KH} =
\frac{\mcloud\vrel}{\fkh\chi^{1/2}\rcloud\sqrt{1+\mach}},
\label{eqn:mlr_kh}
\eq
where $\fkh$ is a free parameter that controls the growth rate of the
KHI, $\chi \equiv \rhocloud/\rhoa$ is the density contrast between the
cloudlets and their ambient medium.
\bq
\dot{M}_\mathrm{c;ev} = 1.8\rcloud\lcloud\dot{\hat{m}}_\mathrm{A},
\label{eqn:mlr_ev}
\eq
where $\dot{\hat{m}}_\mathrm{A}$ is the average mass loss rate per unit
area from evaporation \citep{phewi}.

In a cool ambient medium, thermal evaporation is negligible, while in
a hot ambient medium, where thermal conduction is efficient, we
suppress the mass loss rate from hydrodynamic instabilities by a
factor of $\exp(-\rcloud/\lkh)$. The cloudlet size relative to a
characteristic scale $\lkh$ determines whether or not thermal
conduction is strong enough to suppress the Kelvin-Helmholtz
instability:
\bq
\label{eqn:lambda_kh}
\lkh = \frac{1+\chi}{\chi^{1/2}}\frac{1}{\mach(\mach^2+3)}
\frac{\fS\kappa}{\namb\Ta^{1/2}}
\left(\frac{4\mu\mh}{\gamma^3\kB^3}\right)^{1/2},
\eq
where $\namb$ is the number density of atoms in the ambient medium.

The cloudlets decelerate owing to ram pressure $\pram$ (and
gravitational) forces. The ram pressure in front of a cloudlet is \citep{mc75,
phewi}
\bq
\pram =
\begin{cases}
\pa \left [ \frac{2\gamma}{\gamma+1}\mach^2 -
\frac{\gamma-1}{\gamma+1} \right]
\etas\taus, & \mach > 1;\\
\pa \left( 1 + \frac{\mach^2}{3}\right)^\frac{5}{2}, & \mach \leqslant 1.
\end{cases}
\eq

We decelerate the PhEW particle that represents the cloudlets at the
same rate as the cloudlets:
\bq
\dot{\mathbfit{v}_i} = \frac{\pram\rcloud^2(\vrel/|\vrel|)}{\mcloud} + \mathbfit{g},
\label{eqn:deceleration}
\eq
where $\mathbfit{g}$ is the gravitational field at the particle
location.

In the cosmological simulations, as the cloudlets in a PhEW particle
lose their mass to the ambient medium, the mass of the PhEW particle
decreases accordingly
\bq
\frac{\dot{M}_i}{M_i} = \frac{\dot{\mcloud}}{\mcloud}.
\eq

At the same time, we distribute the lost material to the neighbouring
particles within the smoothing kernel of the PhEW particle in a way that
conserves mass, energy, and metallicity. Here, we
define $\delm$, $\delp$, $\delv$ and $\delE$ as the change of mass,
momentum, velocity, and energy during a time-step $\delt$ for the PhEW particle
as a whole. For each of the
neighbouring particles:
\bq
\delm_j = \frac{\delm_iW_{ij}}
{\sum_j^{\nngb}m_jW_{ij}},
\label{eqn:dm_ngb}
\eq
\bq
\delp_j = \frac{\delp_iW_{ij}}
{\sum_j^{\nngb}m_jW_{ij}},
\label{eqn:dp_ngb}
\eq
\bq
\delv_j = \frac{M_jv_j + \delp_j}{M_j + \delm_j}.
\label{eqn:dv_ngb}
\eq

The metallicity of the neighbouring particle becomes
\bq
Z_j(t+\delt) = \left( M_jZ_j(t) + \frac{\delm_iZ_i(t)W_{ij}}
{\sum_j^{\nngb}m_jW_{ij}} \right) / (M_j + \delm_j).
\label{eqn:dZ_ngb}
\eq

The total amount of heat generated from the ram pressure approximately
equals the net loss of kinetic energy from the PhEW particle and
all its neighbouring particles, i.e.,
\bq
\delE = \delE_{\mathrm{kin},i} + \sum_j^{\nngb}\delE_{\mathrm{kin},j}.
\label{eqn:dE_ram}
\eq

In the limit of strong shocks, a fraction $\chi/(1+\chi)$ of the heat
is advected with the ambient flow while the remaining fraction
$1/(1+\chi)$ of the heat enters the cold cloudlet. Therefore, the
specific energy of a neighbouring particle changes by
\bq
\dot{u}_j = \frac{\left[ M_ju_j + \delm_ju_i +
\left(\frac{\chi\delE}{1+\chi}\right)\frac{W_{ij}}
{\sum_j^{\nngb}m_jW_{ij}} \right]}{(M_j + \delm_j)\delt}.
\label{eqn:du_ngb}
\eq

To prevent neighboring particles from overheating owing to multiple wind particles, we do not allow wind particles to heat any normal gas particle above the post-bowshock temperature, $T_\mathrm{ps}$, calculated for each wind paticle separately according to \citet{phewi}, consistent with the simulations of \citet{bs16}. When multiple wind particles are heating a normal gas particle at the same time, we choose the temperature cap as the maximum $T_\mathrm{ps}$ among all the wind particles. In practice, implementing this temperature cap affects only a very small fraction of the gas particles in the most massive haloes and has nearly no distinguishable effect on galaxy properties such as the galactic stellar masses.

As we evolve the PhEW particles over time, we need to choose the
proper time-step for each particle. To ensure that all the
critical time scales, i.e., the Courant time scale, the deceleration time 
scale, the cloudlet disruption time scale, and the heating time scale, are
resolved, we choose the time step as the minimum of these time scales
multiplied by a parameter $\ft$:
\bq
\delt_i = \ft\,\mathrm{min}(\frac{h_i}{|\vrel|},\,
\frac{\mathrm{max}(|\vrel|, 100\kmss)}{|\dot{\mathbfit{v}}_\mathrm{rel}|},\,
\frac{\mcloud}{\dot{M}_\mathrm{c}},\,
\frac{u_i}{\dot{u}_i}).
\eq
In our simulations, we choose $\ft = 0.2$. When calculating the deceleration time-scale, we use a lower limit for the relative velocity, $\vrel > 100$\kms, to prevent particles that have slowed down from having a time-scale that is too small. Since the cloudlets usually
have small cross sections and cool very efficiently, the deceleration
and heating time-scales are often very long. Therefore, the Courant
condition and the disruption time typically determine the time step of PhEW
particles.

When a PhEW particle has lost a significant fraction, $\flim$, of its
original mass, we remove it from the simulation and distribute its
remaining mass, momentum, and metals among its neighbouring particles in
a single time-step using Equations \ref{eqn:dm_ngb} to
\ref{eqn:du_ngb}. In this paper, we choose $\flim = 0.1$. There is a
very small likelihood that a PhEW particle moves into a galaxy, i.e., the ambient density of the particle becomes higher than the SF density threshold $\rho_\mathrm{SF}$. When
this occurs, the physics that governs the cloudlets is no longer
valid, but the properties of the PhEW particle might be similar to the
surrounding medium. Therefore, if this occurs we transition
the PhEW particle back into a
normal gas particle. In \sect{sec:recouple}, we will explore the option
of converting PhEW particles back into normal gas particles at late
stages in their evolution, i.e., recoupling, instead of removing them.
However, in the end we decided not to allow recoupling 
because it led to potential numerical artefacts, e.g. artificial rapid heating
of the new gas particle.

As the PhEW particles lose their mass to their neighbouring gas
particles, normal gas particles in the simulations can gain
much more wind material than their original mass over time. We found
in our test simulations that over half of the gas particles have
doubled their mass by the end of the simulation ($z=0$) and 5\% of them have
accumulated 10 times or more wind material than their initial mass. The
situation tends to be worse in higher resolution simulations. This
results in a wide range of particle masses at lower redshifts, which
lowers the effective resolution and may cause numerical errors in \textsc{gizmo}.
Therefore, we split any over-massive particle into two smaller
particles once its mass has tripled to prevent a loss of resolution and
other numerical issues, a feature already present in \textsc{gizmo}.

\section{Simulations}
\label{sec:simulation}

\tab{tab:simulations} lists the simulations that we use for this
paper.  In our naming convention ``l'' means $\Lambda$CDM, the first number is
the length of one side of the cubic periodic volume in $h^{-1}$Mpc, and the
number $n$ after the ``n'' indicates that the simulation starts with $n^3$ 
particles of each dark matter and gas.  The table indicates the mass resolution
of each simulation as the mass of one gas particle. Typically we have found
that it requires 128 particles to accurately reproduce galaxy masses when
comparing simulations of different resolutions \citep{finlator06}, but that could
be different for the PhEW model.  We also indicate the spatial resolution
as the comoving Plummer equivalent gravitational softening length $\epsilon$, though the actual form of the softening we use is a cubic spline \citep{hernquist89}.
All simulations use the same initial condition with the following
cosmological parameters:
$\Omega_\mathrm{m}=0.3$, $\Omega_\Lambda=0.7$, $\Omega_{b}=0.045$,
$h=0.7$, and $\sigma_8=0.8$.

To evolve the simulations we use
a version of \textsc{gizmo} \citep{gizmo} that derives from the one used 
by \citet{simba}, except that we launch the galactic winds using the
\citet{huang20a} model and do not include any AGN feedback or any
artificial quenching for massive galaxies. Here we solve the hydrodynamic
equations using the MFM method using the quintic spline kernel with
128 neighbours. We also
calculate cooling using the \textsc{grackle-3.1} package \citep{smith17}
with a \citet{haardt12} UV background and model the
$\mathrm{H_2}$-based star formation rate using the \citet{krumholz11}
recipe to calculate the $H_2$ fraction. We model Type-Ia supernova feedback
and AGB feedback, and explicitly track the evolution of 11 species of metals
as in \citet{simba}.

\begin{table*}
        \centering
        \caption{Simulations}
        \label{tab:simulations}
        \begin{minipage}{160mm}
        \begin{tabular}{lccccc}
                \hline
                Model &
$L_\mathrm{box}[h^{-1}\,\mathrm{Mpc}]$\footnote{Comoving size of the
simulation domain.} & $M_\mathrm{res}[\msolar]$\footnote{The mass
resolution of the simulation, defined as the initial mass of each gas
particle.} & $\epsilon[h^{-1}\,\mathrm{kpc}]$\footnote{The spatial resolution of the simulation, defined as the comoving Plummer equivalent gravitational softening length} & $\mcloud [\msolar]$\footnote{Initial cloudlet mass.} &
$\fkh$\footnote{The parameter that controls mass loss rate from the
Kelvin-Helmholtz instability.} \\
                \hline
                l50n576-phew-m5\footnote{Fiducial simulation.} & 50 &
$1.2\times10^7$ & $0.75$ & $10^5$ & 30 \\
                l50n288-gadget3\footnote{The only simulation performed
with the \textsc{gadget-3} code} & 50 & $9.3\times10^7$ & $1.50$ & - & - \\
                l50n288-phewoff & 50 & $9.3\times10^7$ & $1.50$ & - & - \\
                l25n288-phewoff & 25 & $1.2\times10^7$ & $0.75$ & - & - \\
                l50n288-phew-m4 & 50 & $9.3\times10^7$ & $1.50$ & $10^4$ & 100 \\
                l50n288-phew-m5 & 50 & $9.3\times10^7$ & $1.50$ & $10^5$ & 30 \\
                l25n288-phew-m4 & 25 & $1.2\times10^7$ & $0.75$ & $10^4$ & 100 \\
                l25n288-phew-m5 & 25 & $1.2\times10^7$ & $0.75$ & $10^5$ & 30 \\
                l25n288-m4-fkh30 & 25 & $1.2\times10^7$ & $0.75$ & $10^4$ & 30 \\
                l25n288-m5-fkh100 & 25 & $1.2\times10^7$ & $0.75$ & $10^5$ & 100 \\
                l50n288-phew-m5-rec\footnote{Identical to the
l50n288-phew-m5 simulation except that we allow PhEW particles to
recouple once their mach number drops below 1.0.} & 50 & $9.3\times10^7$ & $1.50$ & $10^5$ & 30 \\
                \hline
        \end{tabular}
\end{minipage}
\end{table*}

Among the \textsc{gizmo} simulations, the l50n288-phewoff simulation
and the l25n288-phewoff simulation do not use the PhEW model for
evolving galactic winds. We use them as baseline models to compare
with the simulations that do use the PhEW model. These two simulations only
differ in their volume size and numerical resolution.

The fiducial simulation for this paper, l50n576-phew-m5, was performed
in a cubic volume with periodic boundary conditions and a comoving
size of $50\,h^{-1}$ Mpc on each side, starting with $2\times 576^3$
particles, an equal amount of gas and dark matter particles, with a
mass resolution of $1.2\times10^7\msolar$ for the gas particles. The
PhEW model is parameterized with a initial cloudlet mass of
$\mcloud = 10^5\msolar$, a Kelvin-Helmholtz coefficient of $\fkh = 30$.
The parameter $\fkh$ reflects a combination of potential effects that could increase the longevity of cold clumps, such as very small-scale radiative cooling, magnetic field effects (e.g. \citet{mccourt15, li19}, though see \citet{cottle20}), and the effects of a bulk outflow by which an entrained clump may experience a lower velocity relative to its surroundings than versus the ambient medium.

In addition, we set the conductivity coefficient $\fS$, to 0.1 for all
PhEW simulations, limiting the strength of thermal conduction to 10\%
of the Spitzer conductivity. Full Spitzer rate conduction seems unlikely to
be realised in nature, as even a weak magnetic field could significantly suppress
conduction perpendicular to the field lines.
Furthermore, by performing further cloud-crushing 
simulations similar to those
in \citet{bs16} we find that amount of conduction is still 
sufficient to suppress the Kelvin-Helmholtz instability.

The rest of the PhEW simulations start with $2\times 288^3$ particles
and vary either in the numerics, e.g., volume size and numerical
resolution, or in the PhEW parameters. Namely, the l50n288-phew-m4 and
the l25n288-phew-m4 simulations have smaller cloudlets with $\mcloud =
10^4\msolar$, which lose mass more quickly, but also use a higher
$\fkh = 100$ to further suppress mass loss in the non-conductive
regime.

The l25n288-m5-fkh100 is identical to the l25n288-phew-m5 simulation, except for
using $\fkh=100$ instead of $\fkh=30$, so that clouds are more resistant to hydrodynamic
instabilities. Similarly, the l25n288-m4-fkh30 is identical to the l25n288-phew-m4 simulation
except for using $\fkh=30$ instead of $\fkh=100$. We introduce these two simulations to understand
to what degree the $\fkh$ parameter affects cloud evolution in PhEW simulations.

Finally, we performed the l50n288-phew-m5-rec simulation to examine an
alternative method for recoupling the PhEW particles. This simulation
is identical to the l50n288-phew-m5 except that it allows a PhEW
particle to recouple as a normal gas particles once its velocity
becomes sub-sonic, while other PhEW simulations forbid recoupling
unless a PhEW particle hits a galaxy. 

For each output from the simulations, first we identify galaxies by
grouping their star particles and gas particles whose density satisfy the star
forming criteria using \textsc{skid} \citep{keres05, oppenheimer10}.
Then we identify the host halo for each galaxy based on a spherical
overdensity (SO) criteria \citep{keres09a}.
We start with the centre of each galaxy and
search for the virial radius $\rvir$ with an enclosed mean density equal to
the virial density \citep{kitayama96}. 
We define the virial mass $\mvir$ as the total mass
within $\rvir$ and the circular velocity as
$\vcirc = (G\mvir/\rvir)^{1/2}$.

\section{PhEW Particles in Simulations}
\label{sec:phewparts}

In this section, we study the properties of PhEW particles
and how they evolve in cosmological simulations.


When a normal gas particle becomes a PhEW particle, we give it a unique
ID and track its properties at every time-step until it either is
removed or recouples according to the criteria from
\sect{sec:recouple}. As the PhEW particles travel in the CGM, we also
keep track of the ambient gas properties as well as the virial mass
$\mvir$ and the virial radius $\rvir$ of its host halo from which they
were launched.
\subsection{Tracking PhEW Particles}
\begin{figure*}
  \includegraphics[width=1.90\columnwidth]{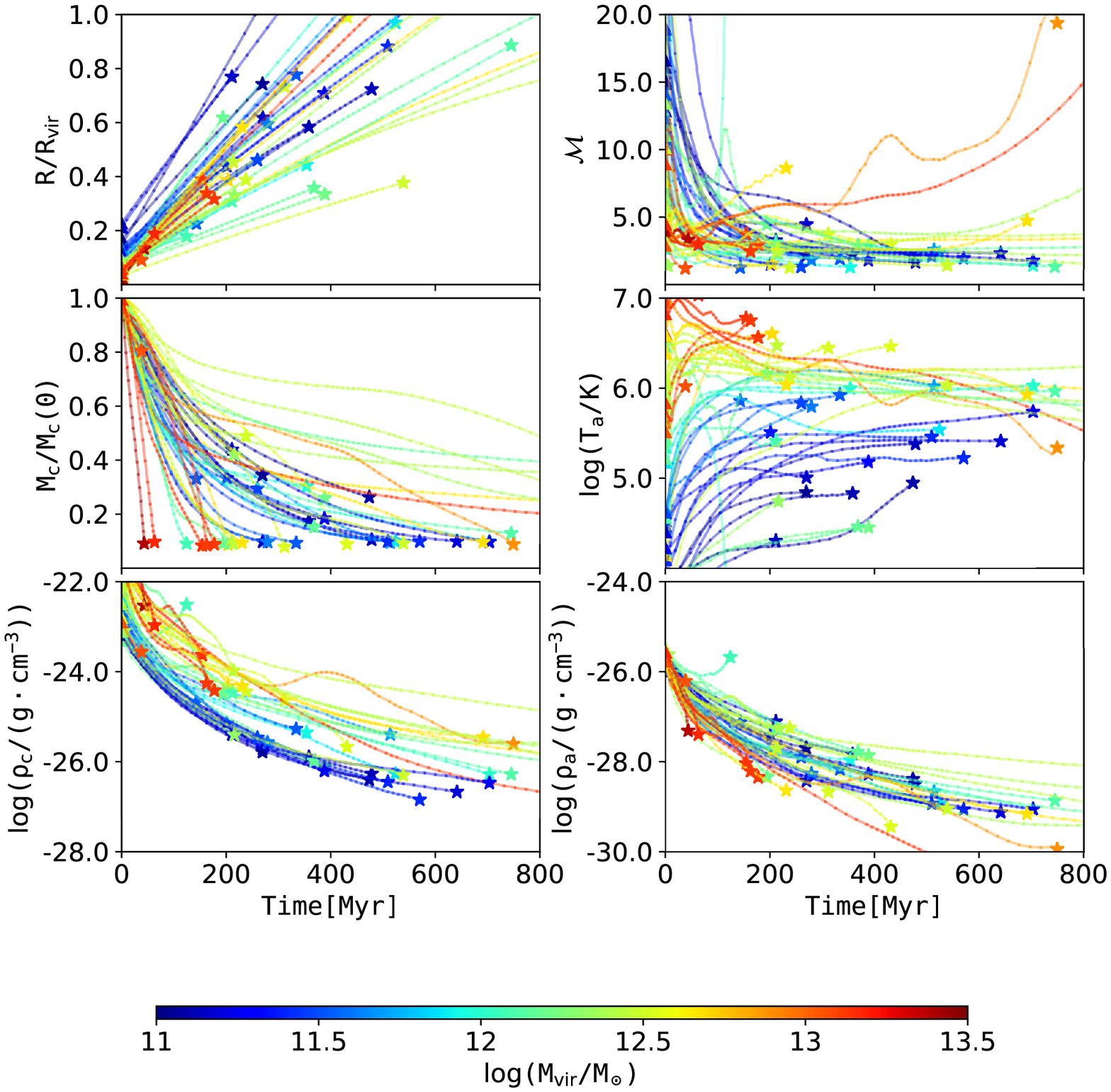}
  \centering
  \caption{The evolution of wind particle properties with time
since they became PhEW particles. We show a random sample of particles
that are launched at $z=1$ from the l50n288-phew-m5 simulation. The
colour of each line corresponds to the mass of its host halo when the
particle was launched, with red tracks coming from massive haloes and
blue tracks from low mass haloes. From top to bottom, left to right,
the properties being plotted are distance from the halo centre
normalised to the virial radius of the halo, the mach number of the
particle relative to the ambient medium, the remaining fraction of
cloudlet mass, the ambient temperature, the cloudlet density, and the ambient
density. The star indicates when the wind particle is removed, i.e. when it has
0.1 times its initial mass.}
  \label{fig:phewtracks}
\end{figure*}

\fig{fig:phewtracks} shows how the properties of PhEW particles and
their surrounding ambient medium change over time, with the stars showing the disintegration time. We select those 
particles that
are launched shortly after redshift $z = 1.0$ from the l50n288-phew-m5
simulation. We choose $z = 1$ for this and some of the later analyses because there are relatively large amounts of wind particles in various haloes covering a wide range of halo mass, but the results are similar at other redshifts because wind evolution is only dependent on the local physical properties of the ambient gas and is therefore not directly dependent on the redshift.
In general, there are much more winds launched from the
low-mass haloes at any given time, but here we choose to show a sample
of particles whose host haloes are evenly distributed across $\mvir$
to demonstrate different wind behaviours in haloes of different masses.

The top panels show how far the cloudlets can travel (left panel) in
their host haloes and how quickly they slow down relative to the
ambient medium. The middle left panel shows how quickly the cloudlets
lose their mass. Most cloudlets lose mass gradually over time and
survive long enough to travel far from their host galaxies and even
beyond their haloes before they disintegrate. The ram pressure and the
gravity slow down the cloudlets rapidly at the beginning but become
less efficient later on. Most cloudlets remain supersonic with $\mach
\gtrapprox 1$ until disintegration.

Thermal conduction has little effect except in the most massive
haloes, where the hot CGM gas ($\Ta > 10^{6.5}\kelvin$) causes the
cloudlets to evaporate within hundreds of Megayears. The linear factor
$\fS$ controls the strength of thermal conduction, but changing this
parameter slightly does not help the cloudlets survive in the massive haloes,
because of the strong non-linear dependence between the conductive
flux and the ambient temperature.

The bottom panels show the density of the cloudlets, $\rhocloud$
(left panel), and the ambient density, $\rhoa$ (right panel). Both the
cloudlet density and the ambient density decrease with time as a PhEW
particle travels towards the outer regions of its host galaxy. The ambient
density decreases as the radius increases and at the
same time the cloudlets expand in the radial direction to adjust
to the change in ambient pressure. Even so, at later times the
cloudlets are still much denser than the ambient medium because,
firstly, the cloudlets are much cooler than the surrounding CGM gas,
especially in the hot haloes, and secondly, the cloudlets are mostly
still travelling at supersonic speeds and experience a large confining ram
pressure.

\begin{figure}
  \includegraphics[width=0.85\columnwidth]{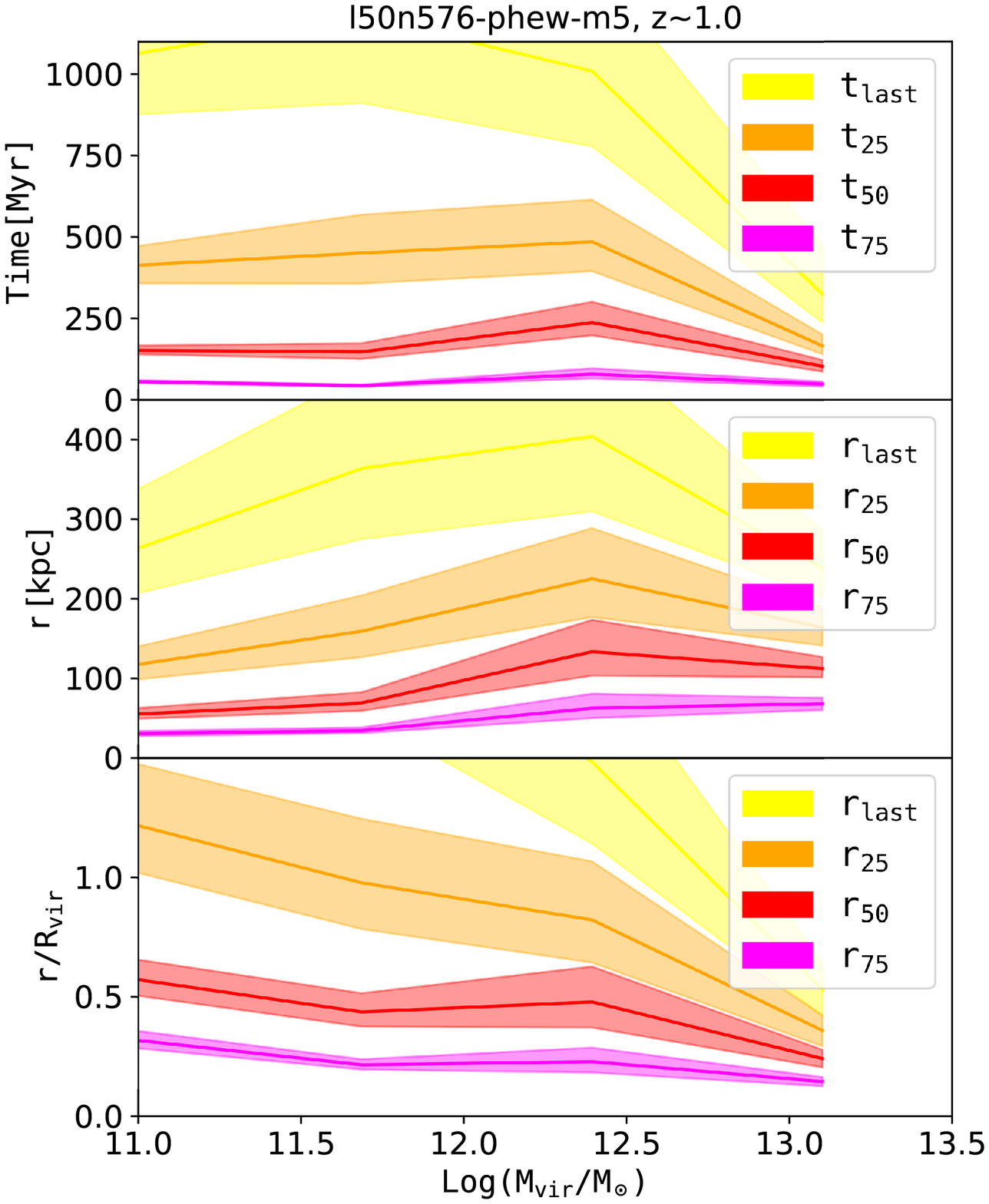}
  \centering
  \caption{How far PhEW particles
travel in galactic haloes of different masses. From top to bottom, the
panels show the time, physical distances and normalised distances that
PhEW particles travel on average at the times when they
have 75\% ($t_{75}$; purple), 50\% ($t_{50}$; red), 25\% ($t_{25}$; orange) of
their original mass
remaining, and at the time when they are removed ($t_{last}$; yellow). The PhEW
particles in this figure are launched at $z=1$ in the l50n576-phew-m5
simulation. The shaded regions indicate the 20th to 80th percentiles.}
  \label{fig:mloss}
\end{figure}

\fig{fig:mloss} shows statistically how far the PhEW particles
travel in haloes of different virial masses. The upper panel and the middle
panel show the average time and the average distances that the PhEW
particles have travelled, respectively. The
different colour lines indicate when and where they have lost 25\%, 50\%, and
75\% of their mass and when they are removed (the disintegration time),
as indicated. In general, the PhEW particles from
massive haloes travel a further distance owing to their higher initial
velocities.  However, the PhEW particles can travel hundreds of kiloparsecs
regardless of their host haloes. 

The lower panel of \fig{fig:mloss} shows the distances normalised to
the virial radius, i.e., $r/\rvir$. In low-mass and intermediate-mass
haloes with $\logmvir < 12.5$, the normalised distances are largely
independent of the halo mass. On average, the PhEW particles are
capable of getting close to the virial radius before they
disintegrate, though they have lost half of their mass when they
reach approximately 0.25 $\rvir$, depositing half of the wind material
within this radius. In massive haloes, conductive
evaporation quickly starts to dominate the mass loss and leads to a rapid
disintegration of the cloudlets. The time and radius at which the
cloudlet disintegrates strongly correlates with the halo virial mass in these
massive haloes as a consequence of the strong dependence between
the evaporation rate and the gas temperature.

To understand the scalings between the PhEW particle mass loss rate and the halo
virial mass, note that in low-mass haloes, hydrodynamic instabilities
such as the Kelvin-Helmholtz instability (KHI) dominate the mass loss from
the cloudlets as the halo gas
is primarily cold \citep{keres05, keres09a}. In haloes above about $10^{11.7}
\msolar$ a halo of hot gas develops from accretion shocks and hence
conductive evaporation becomes increasingly important. In
the KHI dominated haloes, both the wind velocities and the sound speed
of the ambient medium scale as $\mvir^{1/3}$, approximately. 
Compression from ram pressure determines the cloudlet density
so that $\rhocloud \propto \vrel^2$, while the cloudlet radius scales
with the cloudlet density as $\rcloud \propto \rhocloud^{-1/3}$.
Therefore, \eqn{eqn:mlr_kh} suggests that $\dot{M}_\mathrm{c,KH} \sim
\mvir^{2/9}\rhoa^{1/2}$, i.e., the mass loss rate from the KHI depends
only weakly on the halo mass at the same ambient density. In practise,
the PhEW particles last longest in the intermediate-mass haloes.
This is because when the ambient temperature
becomes hot ($\Tvir > 10^6\kelvin$) in these relatively massive
haloes, thermal conduction becomes strong enough to suppress the KHI
but is not strong enough to cause significant conductive evaporation.

In \citet{bs16} as well as many other cloud-crushing simulations, the cloud often disintegrates over short time-scales, e.g., a few cloud-crushing times. The inability of cold clouds to survive is often used as an argument for why outflows cannot explain cold absorption at large distances in massive halos \citep[e.g.][]{zhang17}. In our PhEW simulations, however, the wind particles often survive much longer. The survivability of the clouds in PhEW is controlled by the parameter $\fkh$ for low-mass haloes and the parameter $\fS$ in massive haloes. Both parameters depend on various factors that are not accounted for in idealized simulations such as in \citet{bs16}. For example, clustered star formation might clear out low density regions in the path of the winds making them easier to escape. Tangled magnetic fields suppress both KHI and conduction by an uncertain fraction.
Moreover, the short lifetimes of clouds in the cloud-crushing simulations partly owe to the sometimes large uniform ambient densities, while in cosmological simulations, the ambient density usually decreases as the wind particles move out in the halo.

\begin{figure}
\includegraphics[width=0.85\columnwidth,angle=270]{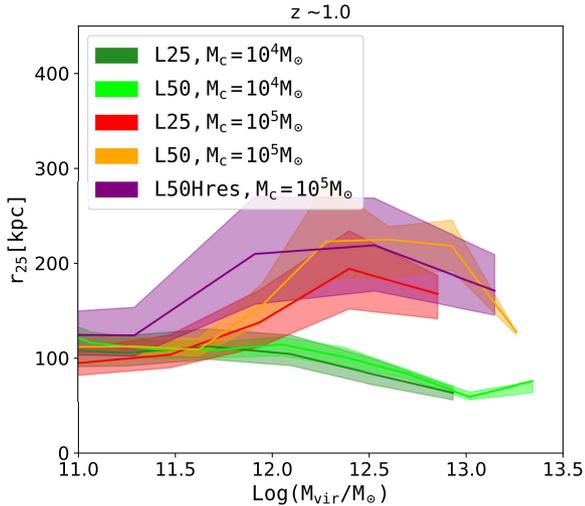}
  \centering
  \caption{The median and the 20th to 80th percentiles of the physical
distances that the PhEW particles travel at the time when they
only have 25\% of their original mass. We compare five
simulations to show the effects of changing the numerical resolution
and the initial cloudlet mass $\mcloud$. The mass evolution of cloudlets is
resolution independent. The smaller mass cloudlets lose mass more rapidly in
haloes where thermal conduction is important.}
  \label{fig:mloss_r25}
\end{figure}

\fig{fig:mloss_r25} compares the physical distances at which the
cloudlets retain only 25\% of their original mass between simulations to
demonstrate how the wind parameters and numerical resolution affect the
distances that the PhEW particles travel. When we assume more massive cloudlets, i.e.,
$\mcloud = 10^5\msolar$, they travel about the same distances in haloes
with $\logmvir < 12$, but survive significantly longer in more massive
haloes than the PhEW particles that have $\mcloud = 10^4\msolar$. In the less
massive haloes where the KHI dominates the mass loss, we compensate for 
the smaller mass of the cloudlets in the $\mcloud=10^4\msolar$ simulations 
by using a $\fkh$ factor that is approximately 3 times larger. In the
more massive haloes, thermal evaporation becomes increasingly
important but the total evaporation time is longer for the more massive
cloudlets, which scale as $\mcloud^{-2/3}$\citep{phewi}.

Importantly, \fig{fig:mloss_r25} also demonstrates that increasing
the mass resolution by a factor of eight, i.e., going from the l50n288-phew-m5 simulation to the l50n576-phew-m5 simulation, or changing the box size by a factor of two, i.e., going from the l25n288-phew-m5 simulations to the l50n576-phew-m5 simulations, does not change $v_\mathrm{25}$ significantly in haloes of any mass, the change being at most 50\%.
This resolution insensitivity of the PhEW model is likely better than other
sub-grid wind algorithm that under-resolve wind - halo
interactions, as a future convergence test of, e.g., $v_\mathrm{25}$, could demonstrate.
For example, \citet{mufasa} find that much of the lack of resolution convergence likely occurs because of poor convergence in wind recycling, owing to wind-ambient interactions being highly resolution dependent.

Since the mass and metals that a PhEW particle loses are mixed into
the ambient medium, \fig{fig:mloss_r25} also suggests how wind
material is distributed within the haloes. In simulations without
the PhEW model, the metal rich wind material is locked to the wind
particles. The material does not mix with the halo gas and
could only fuel future star formation through wind
recycling \citep{oppenheimer10, angles-alcazar17}. With the PhEW model, the
halo gas is constantly enriched by the PhEW particles that pass
through. We, therefore, expect that the structures of cold gas, cooling
and accretion within the gaseous haloes will be very different between
simulations with and without PhEW, as we will show in
\sect{sec:baryons} and \sect{sec:accretion}.


\begin{figure*}
  \includegraphics[width=1.60\columnwidth,angle=270]{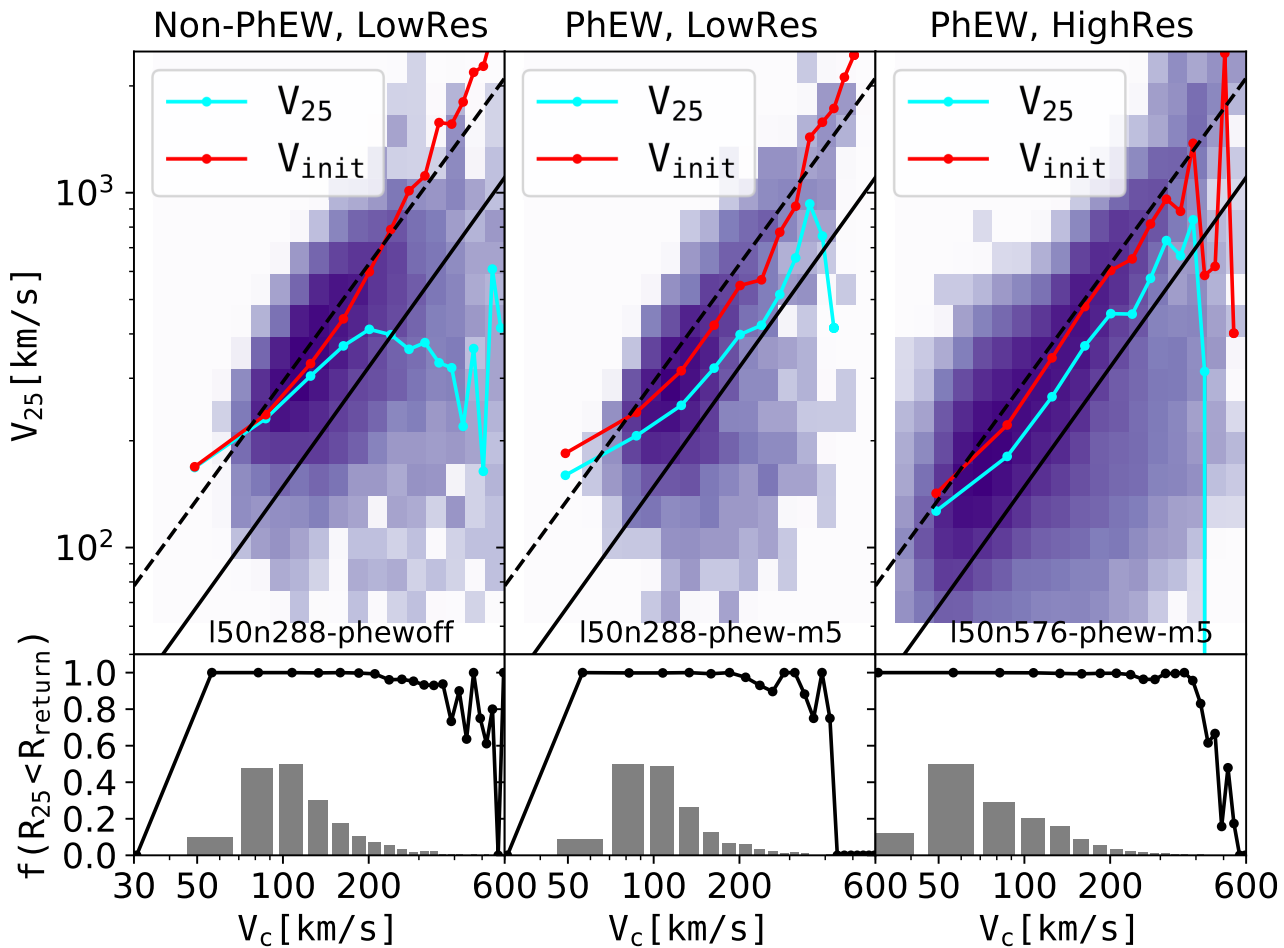}
  \centering
  \caption{\textit{Upper panels: }The relation between the wind
velocities at 0.25 $\rvir$, i.e., $v_\mathrm{25}$, and the halo
circular velocities $\vcirc$. In each panel, the cyan line indicates
the medians of $v_\mathrm{25}$ and the colour scales indicate the
scatter from the relation. The red ($v_{init}$) lines indicate the wind velocity
at launch. The black solid and black dashed lines are
the 50th and the 95th percentiles from \citep{muratov15},
respectively. \textit{Lower panels: }The grey histograms show the
distribution of $\vcirc$ for all selected wind particles. The black
lines show the fraction of wind particles in each $\vcirc$ bin that
were able to reach 0.25 $\rvir$ in their host haloes. From
\textit{left} to \textit{right}, 
the first row shows results
for the \textsc{gizmo} simulation without the PhEW model, and the second
and third rows show results both with the PhEW model but at lower and
higher resolutions, respectively, as labelled.}
  \label{fig:v25vc}
\end{figure*}

\fig{fig:v25vc} shows the scalings between the wind velocities and the
circular velocities, $\vcirc$,
of their host haloes. In each panel,
the colour scale indicates the velocities at 0.25 $\rvir$ and the blue
lines show the medians for each halo mass bin. We compare this scaling
relation from our simulations to the results from high
resolution zoom-in simulations \citep{muratov15}. The solid and dashed
lines in this figure indicate their 50th and 95th percentiles wind
velocities, respectively. The bottom panels show the fraction of wind
particles in each halo mass bin that are fast enough to reach 0.25
$\rvir$.

The wind speed scalings in our simulations align better with the 95th percentile from \citet{muratov15} than their median because of the different ways of measuring the wind speed \citep{huang20a}. In \citet{muratov15}, the wind speed is averaged over all outflowing particles at 0.25 $\rvir$, effectively measuring the bulk motion of all the halo gas at that radius. In our simulations, we average over wind particles only, which should correspond to the fastest outflowing particles in a halo. Therefore, it is more reasonable to compare the wind speed measured in our simulations to the 95th percetile from \citet{muratov15}.


In \citet{huang20a}, we adjusted the wind launch velocities in our SPH
simulations using \textsc{gadget-3} to recover the superlinear scaling
relation from \citet{muratov15} (see the appendix).
Most of the winds in the SPH simulations were then fast enough to
reach 0.25 $\rvir$ regardless of the halo mass. However, in the
\textsc{gizmo} simulation without the PhEW,
even though we use the same wind launch velocities, the
winds slow down more quickly in massive haloes and many of them stop
and turn back before reaching 0.25 $\rvir$ (left panel of \fig{fig:v25vc}).

In the \textsc{gizmo}
simulations with the PhEW model, however, the scaling relation is not only
consistent with \citet{muratov15} but also much less dependent on the
numerical resolution.
In both the SPH and the \textsc{gizmo} simulations without the PhEW,
increasing the numerical resolution tends to slow down winds even more quickly. In practise, we
need to boost the initial wind velocities by a factor of 1.14 in the
SPH simulations when increasing the spatial resolution by a factor of two
(a factor of eight increase in mass resolution) to obtain the
same $v_\mathrm{25} - \vcirc$ scaling relation \citep{huang20a}.
In PhEW simulations, one no longer needs to re-tune
the wind velocities to match the \citet{muratov15} relations at
different resolutions as in the non-PhEW simulations.

Hence, the PhEW model has the advantage that the
dynamics of the wind particles are governed by analytic equations, and thus will be less affected by unresolved or poorly resolved wind-CGM hydrodynamic interactions.

In the PhEW simulations, fewer wind particles reach 0.25 $\rvir$ in
massive halos as they have lost most of their mass before they reach that radius
in these halos. Typically the fraction of wind particles that reach 0.25
$\rvir$ starts dropping significantly above $\vcirc \approx
300$\kms, corresponding to haloes with virial temperatures over 10
million degrees. Clouds in these haloes evaporate on very short time-
scales.

\begin{figure}
  \includegraphics[width=0.85\columnwidth]{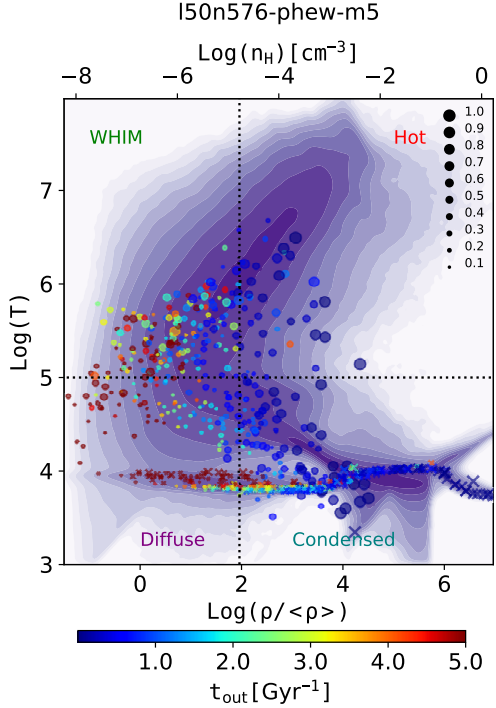}
  \centering
  \caption{The density--temperature gas phase diagram at $z=0.25$ with
PhEW particles over plotted. The background scale indicates the mass distribution
in this phase space. The two dotted lines separate the phase space
into warm-hot gas, hot gas, diffuse intergalactic gas, and 
condensed gas phases. For each PhEW particle, we show its cloudlet properties,
i.e., $\rhocloud$ and $\Tcloud$, as a cross and its ambient
properties, i.e., $\rhoa$ and $\Ta$ as a circle. The size and colour
of the symbol indicates the remaining cloudlet mass fraction and the
time since it became a PhEW particle, i.e., $t_\mathrm{out}$,
respectively. The cloudlets (crosses) are clustered at $\Tcloud \approx
10^4\kelvin$. Older winds tend to have a smaller cloudlet mass and have
travelled further into the lower density regions. In particular, there
are PhEW particles that have survived for more than 5 gigayears. Most of them
are in the most diffuse regions of the Universe.}
  \label{fig:rhot}
\end{figure}

\fig{fig:rhot} shows how the baryonic mass distributes according to
its density and temperature. We only show the result from the
fiducial l50n576-phew-m5 simulation here, but the phase space
distributions are similar between different simulations in this paper.

At $z=0.5,$ PhEW particles have a wide range of properties, such
as $\mcloud$, $\rhocloud$ and the age, $t_\mathrm{out}$, i.e., the age
of the winds since they were launched. The PhEW particles exist in
various halo environments with different ambient properties. A typical
PhEW particle in this diagram travels from the cold, dense interstellar
medium (ISM) and moves across the phase diagram towards lower densities,
with its
cloudlet masses decreasing with time. However, most PhEW particles are at
an equilibrium temperature of around $\Tcloud \approx 10^4\kelvin$
over their lifetimes regardless of the ambient temperature of the surrounding
gas, as a
result of strong metal line cooling inside the cloudlet. This picture is
consistent with high-resolution cloud-crushing simulations. The cloudlet
densities, on the other hand, decrease with time as the cloudlets expand
as they enter more diffuse regions. The cloudlet density is often a few orders 
of magnitude larger than the ambient density owing to compression from
ram pressure and the contrast between the cloudlet temperature and the
ambient temperature.

The PhEW particles often travel at high speed and decelerate slowly
owing to their small cross-sections, so they leave the central regions of their
host haloes very soon after launch and are, therefore, more likely to be
found in the diffuse and the WHIM gas at any given time.

Some PhEW particles
survive for a very long time and can be found in very
diffuse regions of the simulation. Cloudlets in these PhEW particles will have
expanded greatly and become kpc-scale metal rich structures in the
IGM. Most of these long lived wind particles were launched from galactic
haloes with $12.0 < \logmvir < 13.0$, corresponding to a
temperature range where mass loss is minimal because there is still enough
conduction to suppress the Kelvin-Helmholtz instabilities but not enough
to promote rapid conductive evaporation. However, they only represent
a small fraction of the total amount of winds launched from these haloes over time. 
For example, of all the winds
launched from these haloes at $z=1$, less than 10\% have ages over
5 gigayears, and the fraction is much less in other haloes.


\subsection{Wind Recoupling}
\label{sec:recouple}
In our fiducial model, we eliminate PhEW particles from a
simulation when the cloudlet mass drops below a certain fraction $\flim$.
In most circumstances, if a gas particle has been launched as a wind,
it will either remain as a PhEW particle or disintegrate. In this
section, we consider recoupling the PhEW particles that become sub-
sonic, i.e., turning them back into normal gas particles when they
still have more than $\flim$ of their original mass, and we explain
why we decided not to let them recouple in our final PhEW model.

In a test simulation, l50n288-phew-m5-rec, we allow PhEW particles to
recouple when they have slowed down to a sub-sonic speed relative to the
surrounding medium, i.e., $\mach < \machre$, where $\machre$ is a
parameter that we set to 1.0 in our test simulation. When a PhEW
particle recouples, it becomes a normal gas particle that starts
interacting with its neighbouring particles hydrodynamically while
retaining its current mass and velocity.

\begin{figure}
  \includegraphics[width=0.85\columnwidth]{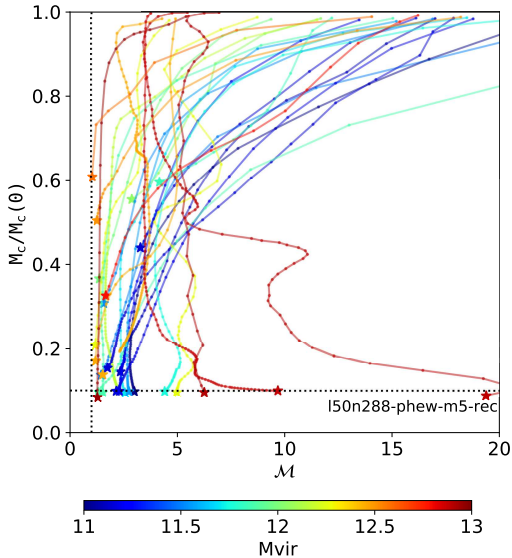}
  \centering
  \caption{The evolution of cloud mass ($M_c/M_c(0)$) as a function of Mach 
  number ($\mach$) for the
PhEW particles launched at $z = 1$ in the l50n288-phew-m5-rec
simulation. The star symbols indicate the last appearance of these
PhEW particles in the simulation. In all the PhEW simulations, we
remove PhEW particles when their mass drops below 10\%, indicated as
the horizontal dotted line. In the l50n288-phew-m5-rec simulation shown here
we also allow PhEW particles to recouple at $\mach < 1.0$,
which is indicated as the vertical dotted line. There is also a small
fraction of PhEW particles that have neither recoupled nor
disintegrated at the end of the simulation. In practise, most
particles have lost enough mass before slowing down to meet the
recoupling criterion.}
  \label{fig:recouple_diagram}
\end{figure}

\fig{fig:recouple_diagram} tracks the evolution of the PhEW particles
before they either recouple or disintegrate in cloud mass--mach number space.
Asymptotically, most PhEW particles
end up in a narrow region of this space where both the
cloudlet mass and the mach number are small. Winds from low-mass haloes
($\logmvir < 12.5$) often start with a high mach number but gradually
slow down and still retain a considerable amount of cloudlet mass at the
time of recoupling. On the other hand, winds from massive haloes
evaporate much more quickly while staying at a nearly constant speed.
\fig{fig:recouple_diagram} also shows a few particles from massive
galaxies with increasing $\mathcal{M}$ at later times as a result of
their surrounding medium becoming cooler as they travel outwards.

Our test simulations also show that there are a small fraction of PhEW
particles that neither recouple nor disappear. Nearly all of them are
launched from intermediate-mass haloes ($\logmvir \approx 12.0$) and
have travelled beyond the virial radius of their host haloes, where
both the mass loss rate and the deceleration rate are very small,
enabling them to stay as PhEW particles for a long time. In our simulations,
we find that around 10\% to 15\% of wind particles launched from these haloes show this behaviour.

In our fiducial model, we remove a PhEW particle once its mass drops
below $\flim=0.1$, indicated as the horizontal dotted line.
Alternatively, one might think of letting it recouple as a normal gas
particle once its mach number becomes low enough, but, we found
that the recoupled particles will often numerically overheat even if
one uses a low mach number for the recoupling criteria (e.g. $\machre
= 1$ as the vertical dotted line in the figure). We experimented by making even 
stricter recoupling criteria. This led to wind particles remaining as PhEW 
particles for longer periods of time before recoupling, during which time
they typically retained a
low temperature.  But, whatever the recoupling criteria, once 
the PhEW particles became ordinary gas particles again, they quickly heated from
about $10^4$ K up to $10^5 \sim 10^7$ K.

This over-heating occurs because recoupling a cold PhEW particle to a hot ambient
medium leads to a sudden change in the estimated density of that
particle.
At the time of recoupling, the density of the particle changes from the cloud density to the ambient density. 
Before recoupling, one uses the density of the clouds to compute the
cooling rate, while after recoupling, one uses the kernel smoothed
ambient density, which is lower than the cloudlet density by a factor of
$\chi \approx \Ta/\Tcloud$. This often leads to an order of magnitude decrease in the density estimation. As a result, the cooling rate drops immediately
at the time of recoupling, enabling the particle to heat up very quickly.

Therefore, we do not allow PhEW particles to recouple in our
simulations owing to this numerical artefact. Our test simulations
show that whether or not we allow recoupling has little effect on 
galaxy properties because a particle usually has lost most of
its mass before meeting the recoupling criteria and this choice, therefore, plays a
limited role in determining galaxy properties. However, this experiment should serve as a
warning to any cosmological simulation that attempts to include the effects of
galactic supernova winds or AGN winds by adding velocities to individual
particles. Such methods might lead to a spurious heating of those particles
compared to a simulation that had orders of magnitude more resolution,
since the evolution of the particle temperature may be sensitive to numerics.

Finally, a PhEW particle may travel into a galaxy, whose further evolution
is beyond the
scope of the PhEW model. Therefore, we let any PhEW particle whose
ambient density is larger than $\rho_\mathrm{SF}$ become a normal
gas particle. However, this affects very few PhEW particles (typically $\ll 0.1\%$).


\section{Galaxy and Halo Properties}
\label{sec:galaxies}
\subsection{The Stellar Masses of Galaxies}
\begin{figure*}
  \includegraphics[width=1.60\columnwidth,angle=270]{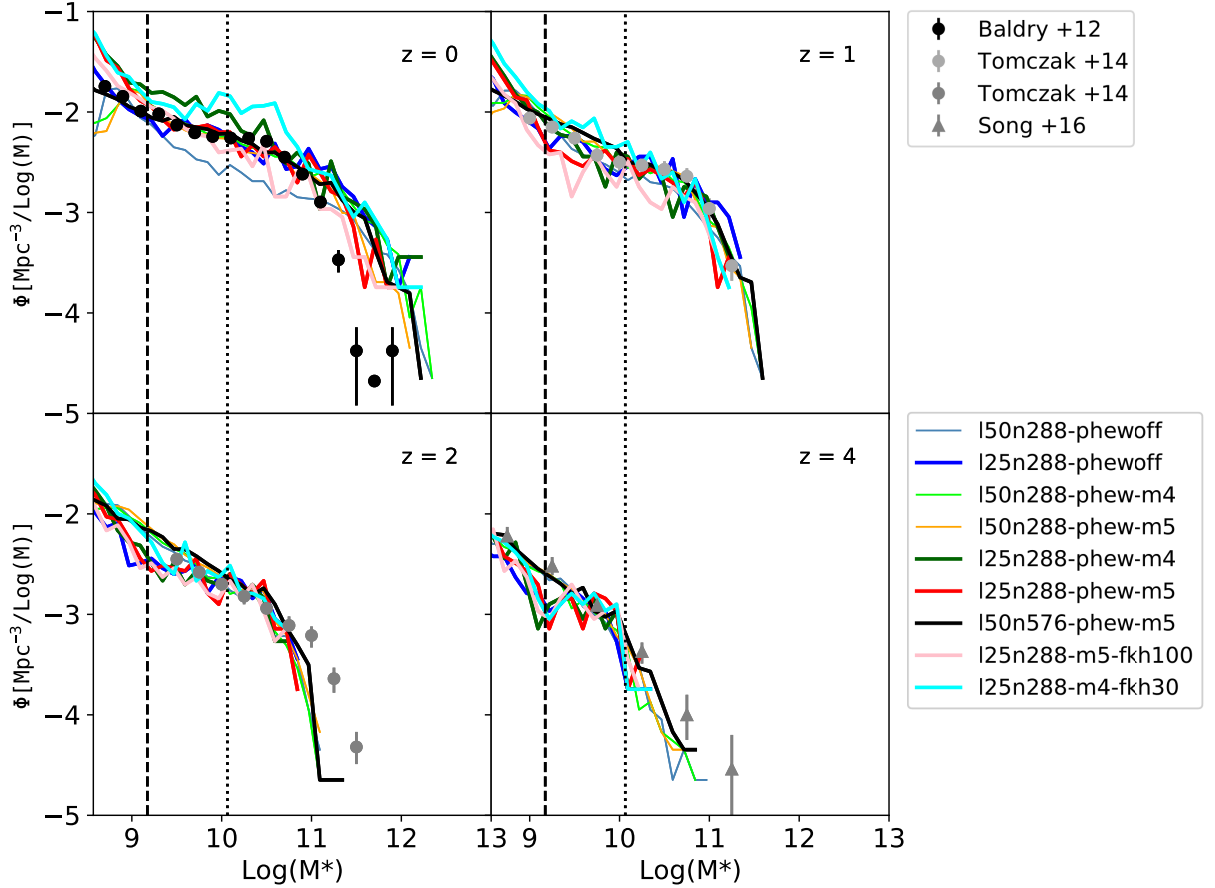}
  \centering
  \caption{Galactic stellar mass functions at $z=0$, $z=1$, $z=2$
and $z=4$. We compare the results from different simulations that are
coloured in accordance with inset key. The two dashed lines correspond to
the two non-PhEW simulations. The bold lines correspond to the simulations
with a higher resolution. See \tab{tab:simulations} for details.
The vertical dotted
and dashed lines indicate the mass resolution limit corresponding to a
total mass of 128 gas particles in the low-resolution and high-resolution
simulations, respectively.}
  \label{fig:gsmfs}
\end{figure*}

\fig{fig:gsmfs} shows the galactic stellar mass functions (GSMFs) at
different redshifts. 
In the simulations without the PhEW, we adjust
wind parameters that control the mass loading factor and the initial
wind velocities so that the GSMFs match observations \citep{huang20a}.
In the simulations with the PhEW, we use the same set of wind
parameters but also adjust the $\fkh$ parameter for each choice of
$\mcloud$. Since a smaller cloudlet loses mass more quickly, we adopt a
larger value of $\fkh$ in the simulations with $\mcloud = 10^4\msolar$.

The GSMFs from the different PhEW simulations with volume sizes and mass
resolutions that vary by a factor of eight are nearly indistinguishable as long as they use the same PhEW parameters such as $\fkh$, 
but the low resolution simulation without the PhEW, i.e.,
l50n288-phewoff, under-produces stars at $z = 0$ relative to other
simulations. The results from the PhEW simulations are much more
resolution independent, regardless of the choice of the cloudlet size.
Surprisingly, in the simulations including the PhEW model, the galactic
stellar mass functions remain viable below the nominal resolution
limit that we established previously of 128 gas particle masses
\citep{finlator06},
down to a mass of 16 gas particle masses as one can see by comparing the lower
resolution l50n288-phew-m5 simulation with the higher resolution 
l50n576-phew-m5 simulation.

On the other hand, changing the PhEW parameters may lead to significant differences in the results. For example, increasing the $\fkh$ parameter help clouds resist hydrodynamic instabilities and enhances their lifetime in low-mass haloes. As a result, the winds from the l25n288-m5-fkh100 simulation generally survive longer than in the l25n288-phew-m5 simulation, resulting in less star formation at $z < 1$. On the contrary, the winds from the l25n288-m4-fkh30 simulation disintegrate faster than in the l25n288-phew-m4 simulation, leading to more recycling and star formation.


\begin{figure}
  \includegraphics[width=0.85\columnwidth,angle=270]{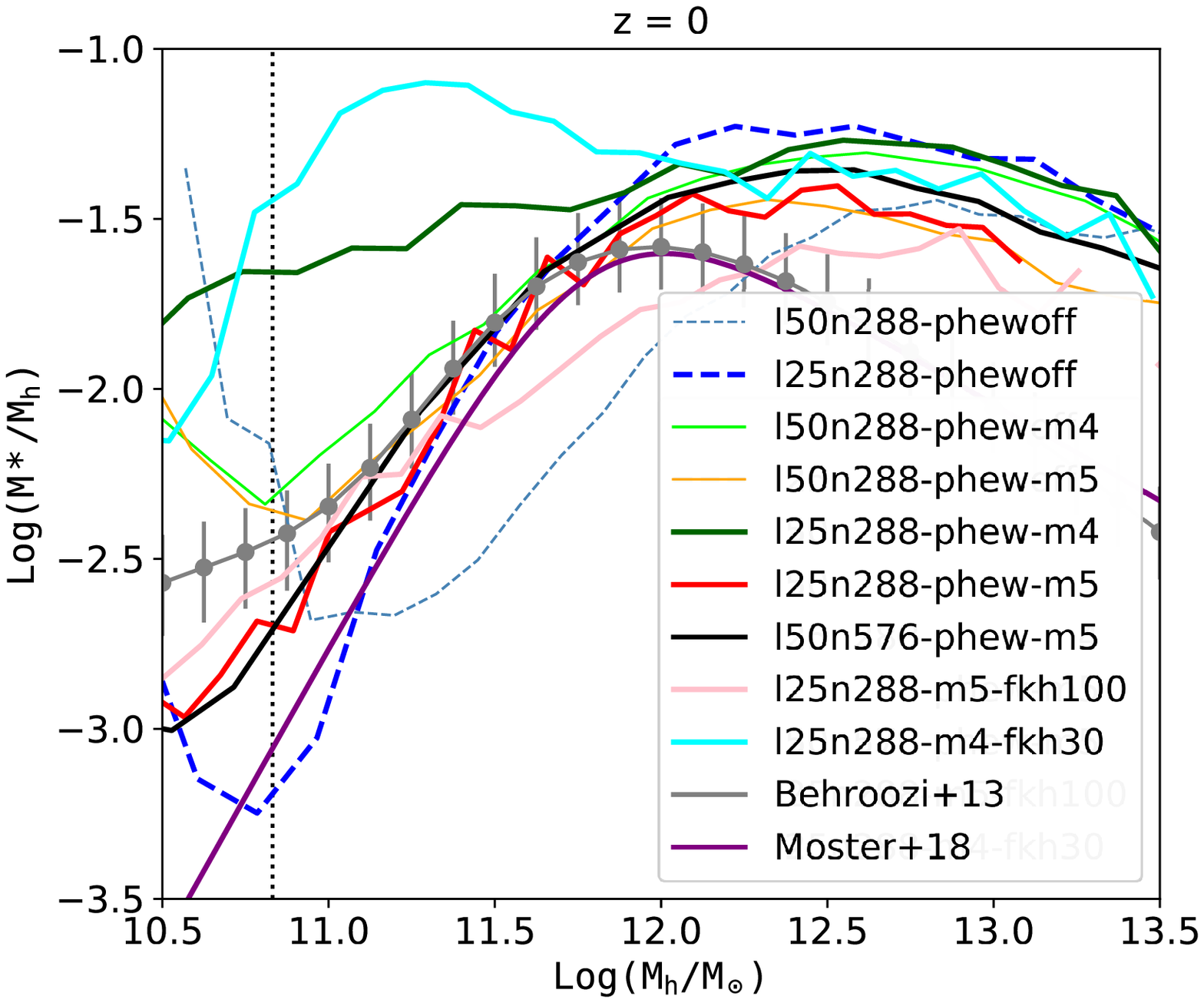}
  \centering
  \caption{Stellar mass - halo mass relations at $z=0$ from the same set of simulations in \fig{fig:gsmfs}. We show the observational results from \citet{behroozi13} (data points with error bars) and \citet{moster18} (purple solid line) for comparison. The vertical dotted line indicate the mass resolution limit corresponding to a total mass of 128 dark matter particles in the low-resolution simulations. The mass resolution limit of the high-resolution simulations is below the plotting limit of this figure.
}
  \label{fig:smhms}
\end{figure}

\fig{fig:smhms} shows a similar result. The $z=0$ stellar mass--halo mass (SMHM) relations between the two simulations without PhEW are significantly different at almost every $M_\mathrm{h}$, while the simulations with PhEW are much more consistent with each other. In particular, the galaxies in low-mass haloes are more massive in the PhEW simulations, in better agreement with observational results from \citep{behroozi13} but worse compared to \citet{moster18}.


\begin{figure*}
  \includegraphics[width=1.40\columnwidth,angle=270]{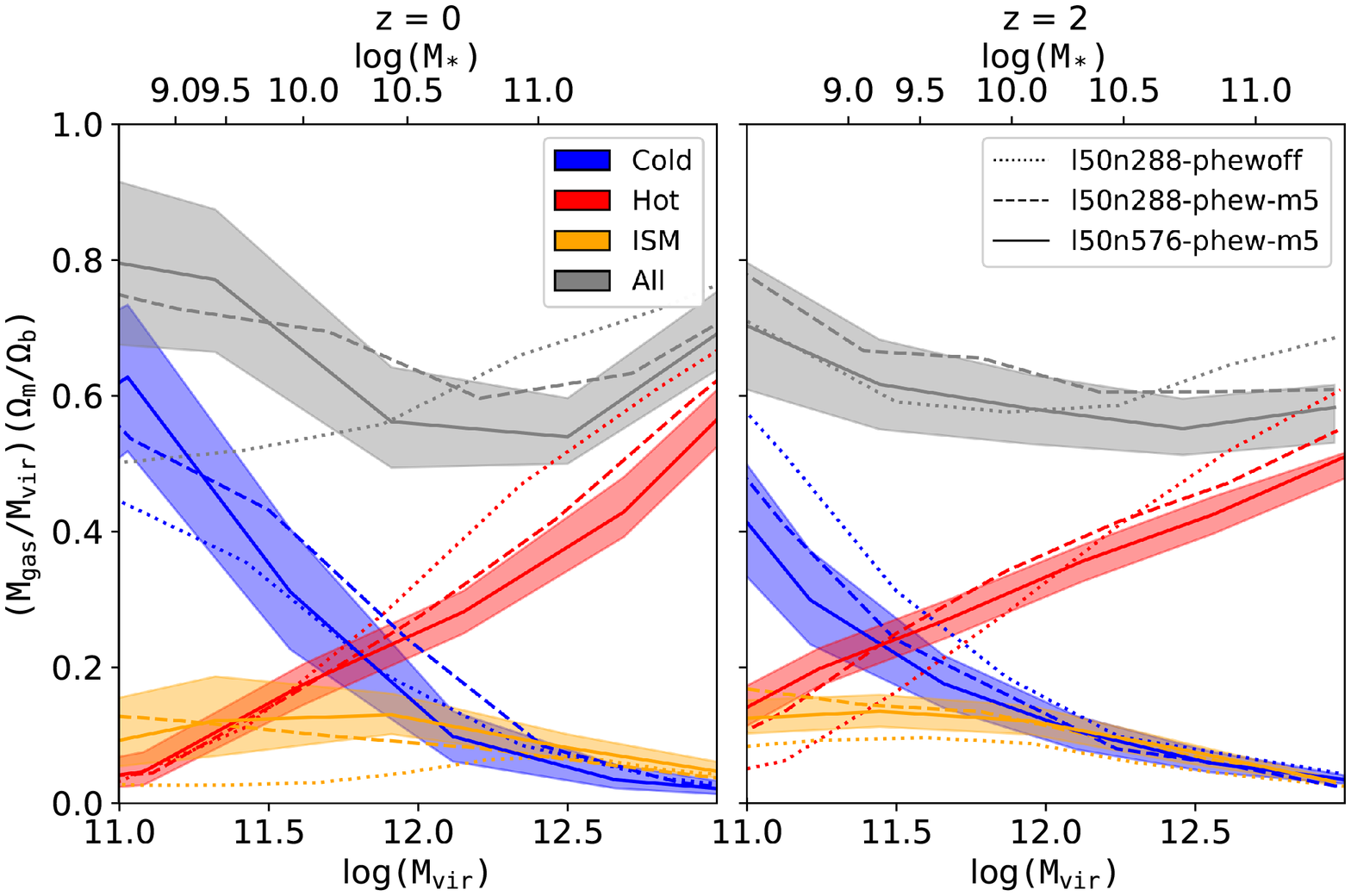}
  \centering
  \caption{The composition of baryons in galactic haloes at $z=0$
(left panel) and $z=2$ (right panel). Each panel shows the mass of
the different components, cold gas (blue), hot gas (red), star-forming
gas (orange) and the total baryons (gas + star, grey), as fractions of
the cosmic mean value with halo mass. The solid lines and the
shaded regions indicate the medians and the 68 percentiles from the
l50n576-phew-m5 simulation. The dashed and the dotted lines are from
the l50n288-phew-m5 and the l50n288-phewoff simulations, respectively.
On the top axis of the panels, we label the average stellar masses of
the central galaxies in corresponding haloes from the l50n576-phew-m5
simulation.}
  \label{fig:halomgas}
\end{figure*}

\subsection{Baryonic Mass Distributions}
\label{sec:baryons}

\fig{fig:halomgas} compares the amount of baryons in different phases
normalised by the cosmic mean baryon fraction. For each central halo,
we separate cold and hot gas with a temperature threshold of
$T_\mathrm{cut} = 10^{5.0}\kelvin$, close to the range of temperatures
where the fractions of many observed ions are very sensitive. In previous
papers, we used a slightly higher threshold of $10^{5.5}\kelvin$
\citep{keres05, huang20a}, but the difference between these values has little effect
on the cold gas fraction. In addition to cold and hot halo gas, we
also show the mass of star-forming gas in the galaxies and the total
baryonic mass including stars.

The fraction of hot gas in galactic haloes depends strongly on the
halo mass but is less sensitive to the different numerical models. The
halo gas switches from mostly cold to mostly hot at around $\logmvir =
12.0$. This threshold is somewhat higher than in \citet{keres05}, owing to the inclusion of metal line cooling \citep{gabor12}.
In general, the different wind models do not significantly
affect the amount of hot gas in galactic haloes of any given mass.
This is also true when we change the PhEW parameters and the numerical
resolution.

The baryon fraction within haloes is less than 75\% of the cosmic
mean baryon fraction in most haloes, regardless of which code or which
wind model we use. This depletion of halo baryons is the result of winds that escape from haloes
over time. The PhEW simulations have a similar fraction of wind
material escaping from haloes as the non-PhEW simulations. The escape
fractions are also similar at $z=0$ and $z=2$, indicating that most of
the winds escape at higher redshifts.



\begin{figure*}
  \includegraphics[width=1.60\columnwidth,angle=270]{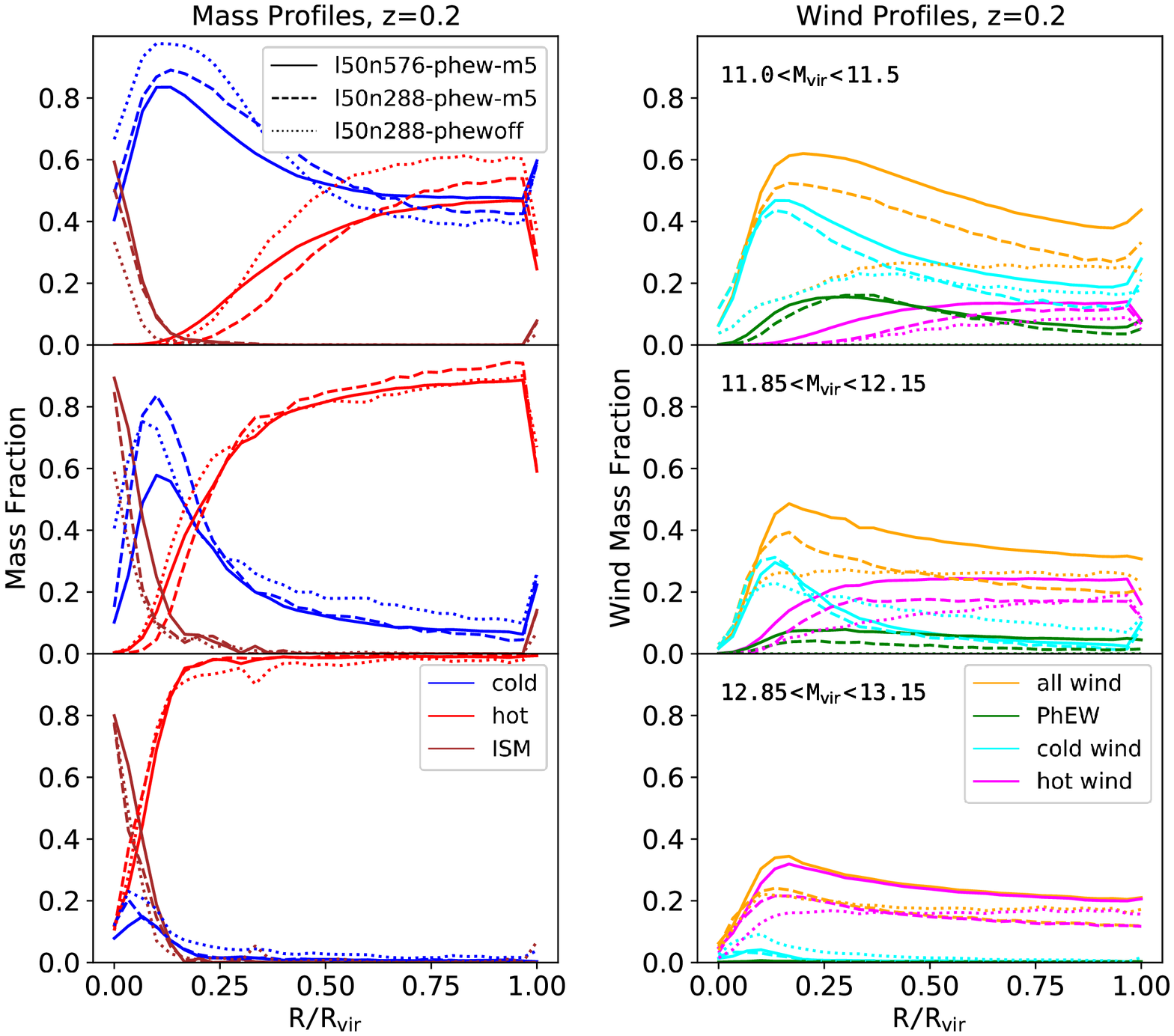}
  \centering
  \caption{The average radial profiles in galactic haloes selected by
halo mass at $z=0.25$. The \textit{left} panels show total baryon mass
profiles including winds. In each radial bin, we separate the baryons
into cold gas (blue), hot gas (red) and star-forming gas (brown) as
described in the text. The \textit{right} panels show the total
fraction of wind material (orange) as a function of radius. We also
separate the wind material into cold wind (cyan), hot wind (magenta)
and PhEW particles (green, only existing for the PhEW simulations).
From \textit{top} to \textit{bottom} shows averaged profiles for low-
mass, intermediate-mass and massive haloes, with their corresponding
mass ranges indicated in the right panels. The dashed and solid lines
compare results from the l50n288-phewoff and the l50n288-phew-m5
simulations, respectively.}
  \label{fig:radmprof}
\end{figure*}

\fig{fig:radmprof} shows how wind particles affect the radial
distribution of baryons and wind material in galactic haloes with
different $\mvir$. In the left panels, we calculate the mass fraction
of the cold, hot and star-forming gas and wind particles as a
function of the normalised radius $r/\rvir$ for each halo and average
over all central haloes within the same $\mvir$ range. We define star
forming gas as dense gas particles residing in galaxies that satisfy
the star-forming criteria that we use during the simulations,
and we separate cold halo gas from hot
halo gas based on the temperature criterion $T < 10^{5.0}\kelvin$.
Consistent with \fig{fig:halomgas}, the fraction of cold and hot gas
as a function of halo radius is similar between non-PhEW and PhEW
simulations, especially in the intermediate and massive haloes.

In the right panels of \fig{fig:radmprof}, we show the fraction of
wind material in each radial bin. For the non-PhEW simulation, we
define wind material as gas particles that were once launched as winds, and
we further separate them into cold winds or hot winds based on their current gas
temperature. These particles often have distinct properties from their
neighbouring normal gas particles, e.g., a higher metallicity as they
were enriched in the galaxies before becoming winds. In the PhEW
simulations, wind particles mostly disintegrate after losing most of
their mass to the neighbouring particles, so that the wind material is 
well mixed with the halo gas. For each gas particle, a fraction of its
mass has come from a PhEW particle. Therefore, we define the
wind fraction in a gas particle as the fraction of the material that used to
be in a PhEW particle. In addition, the currently surviving PhEW particles
make up a small fraction of the total baryonic mass in the low-mass
haloes in the PhEW simulation. This PhEW fraction becomes negligible
in more massive haloes because PhEW particles launch less frequently
and travel faster through these haloes.

Even though the definition of wind material differs between the
non-PhEW and the PhEW simulations, its radial distribution 
between these simulations is very similar in the
intermediate and massive haloes, with a total wind fraction of
approximately 30\% and 20\% at most radii, respectively. 
However, in the low-mass haloes, the wind fraction in the PhEW
simulations is considerably higher. The wind fraction in the PhEW
simulation also increases towards smaller radii until the innermost
region where star-forming gas dominates, while the wind fraction in
the non-PhEW simulation declines within 0.5 $\rvir$ {because typical
wind particles travel quickly and hence spend little time near the galaxy.

The fraction of hot wind material is also very similar between the
two simulations, but the origin of the hot wind differs.
In the non-PhEW simulation, the hot wind material is simply
high velocity wind particles that heat up to the ambient temperature.
In the PhEW simulation, on the other hand, it mostly made up of
cloudlets that evaporate and mix into the hot ambient medium.

Although not shown here, the profiles from the
l50n288-phew-m4 simulation are nearly identical to the l50n288-phew-m5
simulation, except that it has a slightly higher fraction of PhEW
particles in the low-mass haloes.

\subsection{Gas Metallicities}

\begin{figure*}
  \includegraphics[width=1.60\columnwidth,angle=270]{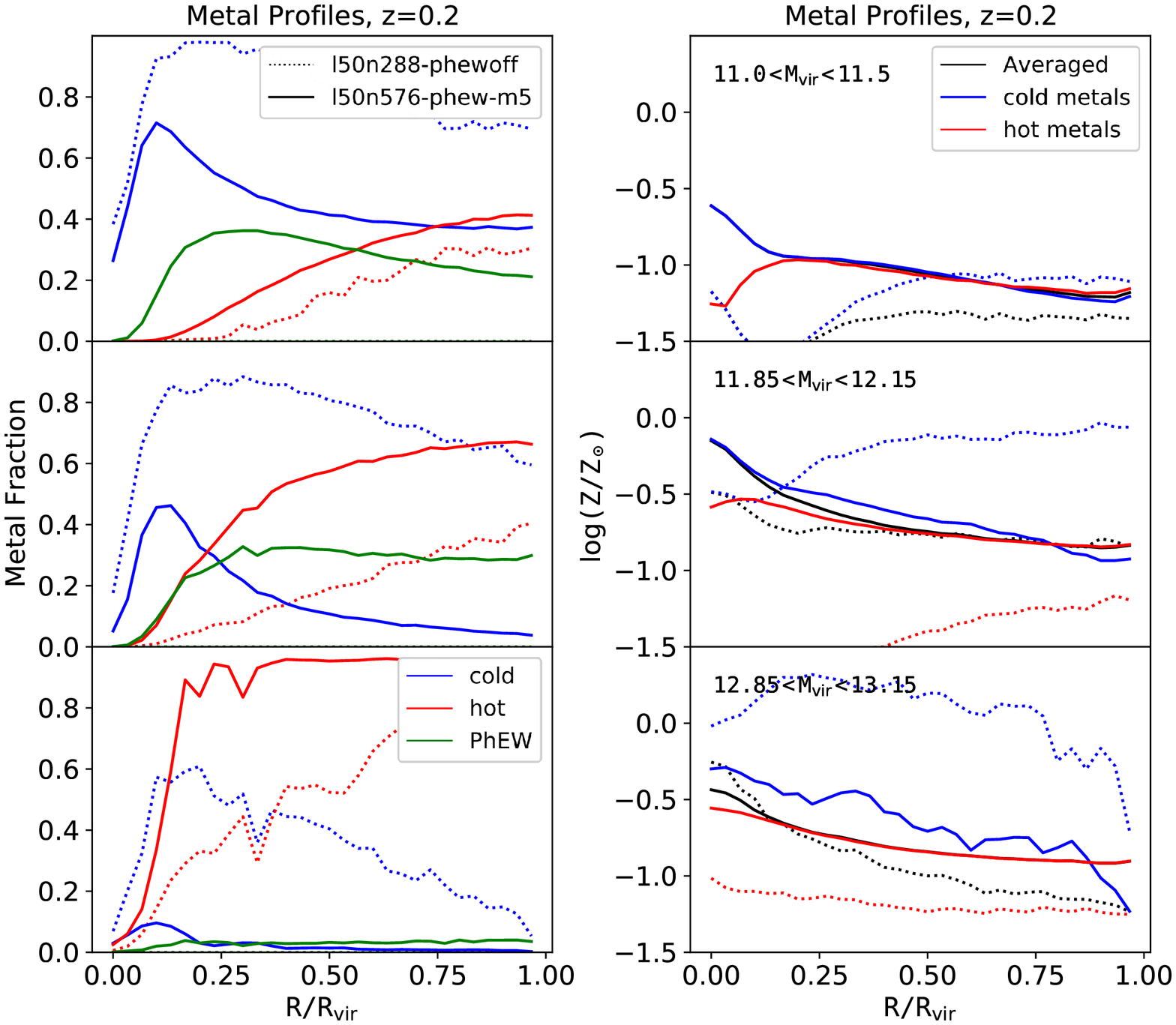}
  \centering
  \caption{Similar to \fig{fig:radmprof}, this figure shows how metals
are distributed in the different gas phases. The \textit{left panels} show the
fraction of metals in cold gas (blue), hot gas (red) and the PhEW
particles (only applying to the PhEW simulation) as a function of
radius. Note here we exclude the ISM gas, which dominates at inner radii, from the plot. The \textit{right panels} shows the metallicity of cold gas (blue), hot gas (hot) and
the average of the two in solar units as a function of radius. Here we use
$Z_\odot = 0.0122$ from \citet{asplund09}.}
  \label{fig:radzprof}
\end{figure*}

The PhEW particles lose not only mass but also metals to the
surrounding medium so the radial distribution of metals also changes
significantly with the PhEW model. In the left panels of \fig{fig:radzprof},
we study the fraction of metals in each of the gas phases as a
function of halo radius. For the non-PhEW simulation, most of the
metals are locked in former wind particles and there is no exchange of
metals between the former wind particles and normal gas particles.
Therefore, one can interpret the cold and hot fraction in the non-PhEW
simulation as the fraction of former wind particles that are currently
cold or hot.

In the PhEW simulation, $20\%$ to $40\%$ of the metals are in the
PhEW particles at most radii in low-mass and intermediate-mass haloes
as they survive longer in haloes of these masses. In addition, the fraction of
metals in the hot gas tends to increase with radius and halo mass. In
the non-PhEW simulation, the majority of metals are in the former wind
particles that remain mostly cold. Even in the most massive haloes
($\logmvir \approx 13.0$), a considerable amount of metals are in the
cold wind particles that scatter over all radii. The former wind particles could remain mostly cold surrounded by hot halo gas particles in part because of their high metallicities and in part because the interactions of single particles with the ambient background are not accurately modeled.
In contrast, almost all metals are 
hot in the PhEW simulation owing to the absence of
cold gas in these massive haloes and the ability to share metals.

In the right panels of \fig{fig:radzprof}, we calculate the metallicity as
the ratio of the total mass of metals and the total gas mass for each phase
separately. Even though the averaged metallicity profiles are similar
between the non-PhEW and the PhEW simulations, the metallicities of
the cold and hot gas alone are drastically different. In the PhEW
simulations, the metallicities of the cold and hot gas are comparable.
However, in the non-PhEW simulations the metallicity of the cold gas
is significantly higher, since the metal-rich former wind particles
are on average colder than the ambient gas particles owing to more
efficient metal cooling, resulting in disproportionally more metals in
the cold gas.

In the most massive haloes, the metallicity of the hot gas in the PhEW
simulation is much closer to the observed universal value of 1/3
$Z_\odot$ for hot cluster gas at low redshifts \citep[e.g.][]{degrandi01, leccardi08, molendi16}, while the metallicity in the non-PhEW
simulation is much lower than this observed value. Although the halos considered here are low-mass groups rather than clusters, a comparison of cluster metallicity profiles can be undertaken in the future.


\begin{figure}
  \includegraphics[width=0.85\columnwidth]{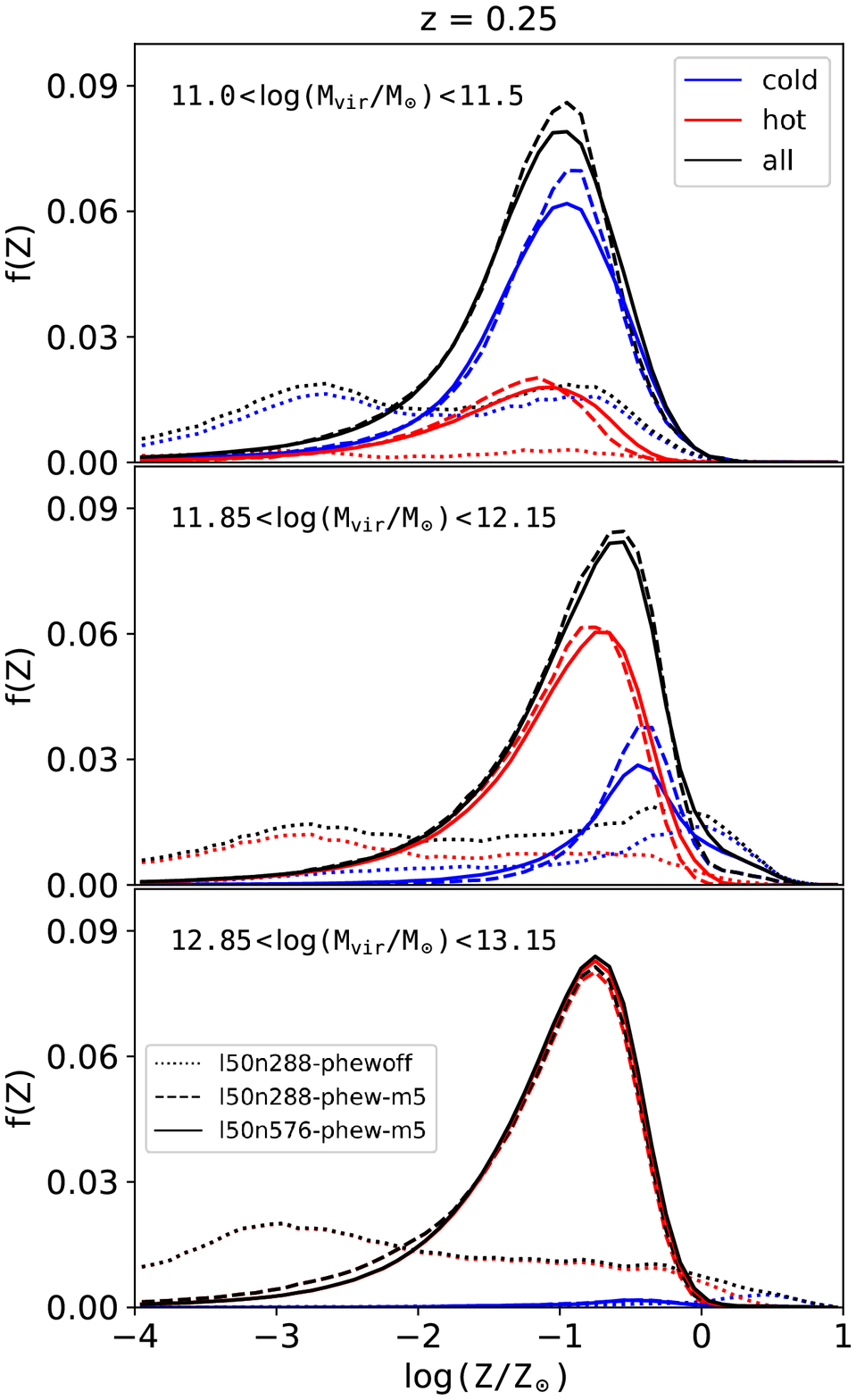}
  \centering
  \caption{Mass weighted histograms of metallicities in low-mass 
(\textit{upper panel}), intermediate-mass (\textit{middle panel}) and 
massive (\textit{bottom panel}) haloes at $z=0.25$. We compute the histograms
using gas particles within the virial radius and exclude
particles that are star forming or that are from the innermost radii 
($r/\rvir < 0.1$). We
normalise the histograms by the total mass of the selected particles. We also
show the histograms for the cold (blue) and hot (red) gas separately, using a
temperature cut of $\log(T/\kelvin) = 5.0$. The dotted, dashed, and solid lines
correspond to the l50n288-phewoff, l50n288-phew-m5, and the l50n576-phew-m5 
simulations, respectively. A considerable fraction ($>40\%$) of particles
in the l50n288-phewoff simulation have metallicities below the plotting
limit of the figure.}
  \label{fig:metalhist}
\end{figure}

\fig{fig:metalhist} compares the metallicity distributions in the halo
gas between the non-PhEW and the PhEW simulations. The two PhEW simulations,
l50n288-phew-m5 and l50n576-phew-m5, have very similar distributions
regardless of the different numerical resolutions, which is another example
of the robustness of the PhEW model. Gas metallicities in the
PhEW simulations
strongly peak at between $-1<\log(Z/Z_\odot)<0$ for all of the three halo mass
bins. The metallicities in the cold and hot gas have similar shapes except
for a slight shift in the peaks. The non-PhEW simulation, on the other hand,
has much broader overall metallicity distributions. Over $40\%$ of particles
in the non-PhEW simulation even have metallicities below the plotting limit,
$\log(Z/Z_\odot) = -4$, of \fig{fig:metalhist}, while the fractions in the two
PhEW simulations are negligible.
The non-PhEW simulation also shows a clear
bimodality in the metallicity distributions in both the low-mass
and intermediate-mass haloes. In the intermediate-mass haloes, the metal-rich
peak at $\log(Z/Z_\odot) \approx -0.5$ is dominated by cold gas while the
metal-poor peak at $\log(Z/Z_\odot) \approx -3.0$ is dominated by hot gas.

In the non-PhEW simulations, the bimodal distribution of metals in galactic
haloes results from different metallicities in the inflowing and outflowing
material \citep{ford14, hafen17} and the inability of these different materials
to mix. This inability to mix metals is present in any hydrodynamical simulation where mass exchange between fluid elements is not allowed (e.g. Gizmo-MFM, SPH, and in contrast to e.g. AREPO and all Eulerian grid simulations) and even in simulations that attempt to
evolve the winds numerically \citep[e.g.][]{schaye15, fire2}
since the mixing occurs on scales that cannot be 
resolved in those simulations.

Many recent simulations \citep{shen10, brook12, brook14, su17, tremmel17, rennehan21}
have implemented sub-grid models to allow metal diffusion 
following the \citet{smagorinsky63} turbulence model. Results from these
simulations suggest metal diffusion plays important roles in the chemical
evolution of dwarf galaxies \citep{pilkington12, revaz16, hirai17, escala18}. However, the physics underlying these models are very different from those in PhEW model and it is not clear if the impact of the PhEW on distributing metals would be similar to the sub-grid diffusion models adopted in these simulations or not.

In the PhEW simulations, metal
mixing can occur naturally between the outflowing wind particles and the ambient medium,
which effectively erases the bimodality. Observationally, whether or not a
bimodality exists at low redshifts is still uncertain
\citep{lehner13, prochaska17}. Future quasar absorption line observations
could potentially distinguish different models of how metals
mix in the CGM. Furthermore, although these results hint at major differences
in the predicted metallicity distributions in the different wind models,
our cold-hot temperature split is only a crude
attempt to make contact with the observations. To make more robust comparisons
with the observations of gas metallicities requires simulated quasar 
absorption lines, which we will pursue in future work.

Despite the fact that the PhEW model drastically changes the distribution of metals
in the CGM, it hardly affects the metallicities of the star-forming
gas in the simulated galaxies. We find that the mass-metallicity
relation in the simulations without PhEW are nearly identical
to that in the simulations with PhEW.

\subsection{Gas Accretion}
\label{sec:accretion}
The gas that ultimately forms stars in galaxies had different thermal
histories before accretion. Galaxies may acquire gas through cold
accretion, hot accretion and wind re-accretion and their relative
importance strongly depends on the galaxy or dark halo mass \citep{keres05,
keres09a, oppenheimer10}. In simulations without PhEW, one often
defines wind re-accretion as the accretion of former wind particles.
The definition is clear since the wind particles do not mix with other
particles.

In simulations with the PhEW, the mass of a normal gas particle can
grow by a considerable amount owing to material lost from neighbouring
PhEW particles before the gas particle accretes onto a galaxy. We
define the wind mass, i.e, $M_\mathrm{w}$, of a gas particle as the mass
that came from PhEW particles. At the time when a gas particle
accretes, we count its wind mass as wind re-accretion and the rest of
its mass as pristine accretion. In non-PhEW simulations, since there
is no mass exchange between normal gas particles and wind particles,
we define pristine accretion as the accretion of a gas particle that
has never accreted onto any galaxy before, and wind re-accretion as
the accretion of particles that used to be winds. This definition for
the non-PhEW simulations is identical to our previous papers
\citep{oppenheimer08, huang20a}.

For each PhEW particle we note the time it was launched, $t_w$, and the velocity
dispersion, $\sigma_w$, of the galaxy that launches it in a wind. Therefore,
whenever a gas particle acquires some mass from a PhEW
particle, we can calculate the mass-weighted average
wind launch time, $\bar{t}_w$, and the mass-weighted velocity dispersion
of the wind launching galaxy $\bar{\sigma}_w$ for the wind component of the
gas particle. We
find that the variances for $t_w$ are usually large while the
variances for $\siggal$ are usually small, indicating that a gas
particle typically accumulates its wind material over a long period of time but
mostly from winds launched from the same halo.


\begin{figure}
  \includegraphics[width=0.85\columnwidth]{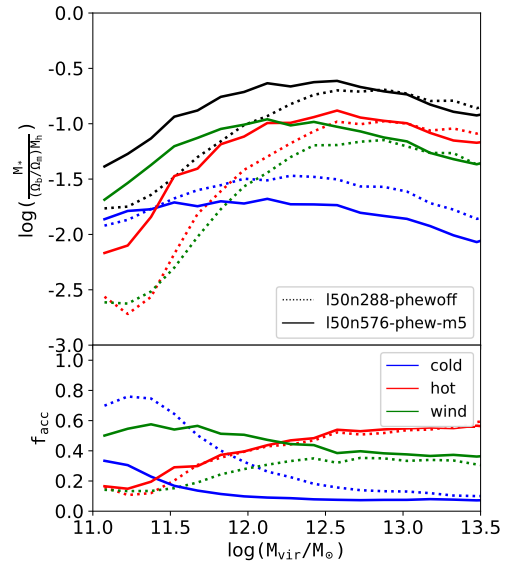}
  \centering
  \caption{
The relative importance of different accretion types
for galaxies of different mass. In the \textit{upper panel}, we
calculate for galaxies at $z = 0$ how much of their total stellar
masses (black) came from cold accretion (blue), hot accretion (red) or
wind re-accretion (green). The \textit{lower panel} shows their relative
fractions. In each panel, the dotted lines are from the
l50n288-phewoff simulation, which defines cold accretion, hot
accretion and wind re-accretion similar to \citet{oppenheimer10}.
The solid lines are from the l50n576-phew-m5 simulation,
using the definitions described in the text. The results from the l50n288-phew-m5 simulation are similar to the l50n576-phew-m5, so we do not plot them on this figure for a cleaner visualization.}
  \label{fig:acc}
\end{figure}

In \fig{fig:acc}, we look at the thermal histories of gas particles
before they accreted onto galaxies and formed stars. For each galaxy
at $z=0$, we consider all star particles of each galaxy, and for those
that accreted onto the galaxy as gas, determine whether they were cold
accretion, hot
accretion or wind re-accretion. The accretion event associated with a star particle is the last time that a gas particle accretes onto a galaxy before becoming a star particle, when its density rises above the star formation density threshold $\rho_\mathrm{SF}$. In a non-PhEW simulation, each
accretion event can only be one of the three types, while in a PhEW
simulation, each accretion event can be a mix of wind re-accretion and
either of the two other types. In this analysis, we separate cold
accretion from hot accretion using a temperature cut of
$T_\mathrm{max} = 10^{5.5}\kelvin$, where $T_\mathrm{max}$ is the maximum
temperature that the gas particle reached before accretion.

In the non-PhEW simulation, l50n288-phewoff, cold accretion dominates
galaxies in small haloes with $\logmvir < 12.0$ while hot accretion
dominates in massive haloes. Hot accretion might become less important
if one were to include AGN feedback to efficiently suppress cooling flows in
the most massive haloes. The fraction of wind-re-accretion is
negligible in very low-mass haloes and grows with halo mass. The
importance of wind re-accretion is sensitive to wind model
implementations \citep{huang20a}. For example, the initial wind speed
determines whether or not wind particles can escape haloes of
different masses and controls the amount and the time-scale of wind
recycling.

Compared to the non-PhEW simulation, the main difference in the PhEW
simulation is a significant increase of wind re-accretion in low-mass
haloes, making it the dominant mode of accretion in these haloes.
Adding this recycled wind material leads to as much as 0.5 dex
increase in the low-mass end of the stellar mass - halo mass relation,
which now matches the empirical results from \citet{behroozi13} as shown in \fig{fig:smhms}.
Although not shown here, the results from the low resolution l50n288-phew-m5 simulation are very similar to the l50n576-phew-m5 simulation.

Most winds can easily escape the low-mass haloes in both the PhEW and
the non-PhEW simulations. However, a PhEW particle loses its mass as it travels
outwards to the cold gas that it passes through and that gas can later accrete 
unto the galaxy. A
star-forming galaxy in a low-mass halo usually has strong winds
resulting in a large amount of wind material being mixed with the accreting
gas and, therefore, increases the amount of wind re-accretion in a PhEW
simulation.

The total amount of cold accretion is similar between the non-PhEW
and the PhEW simulations, but the total amount of hot accretion also
significantly increases in the PhEW simulations, possibly because the
hot gas is more metal rich owing to the mixing with outflowing wind material.

The two PhEW simulations with different resolutions are much more
similar. The l50n576-phew-m5 simulations have slightly more hot
accretion and wind re-accretion but less cold accretion, but the total
amounts of gas accretion in these two simulations are comparable,
again demonstrating the robustness of the PhEW model to numerical
resolution.

\begin{figure}
  \includegraphics[width=0.85\columnwidth]{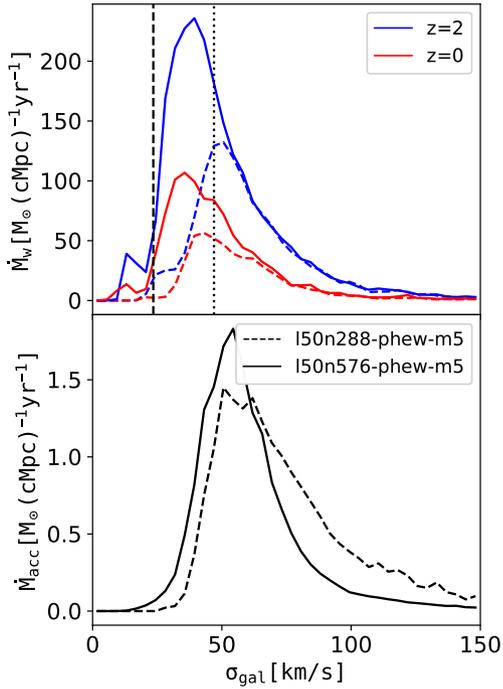}
  \centering
  \caption{In the upper panel, we show from which haloes (binned by their
velocity dispersions) the winds are launched at z = 2 (blue) and at z = 0 (red).
The l50n576-phew-m5 simulation (solid lines) launches much more wind in low-mass
haloes, which are unresolved in the l50n288-phew-m5 simulation 
(dashed lines), at
both redshifts. In the lower panel, we show the mean $\siggal$ of all wind
material that accreted unto the galaxies that are resolved in both simulations within the last gigayear in these simulations.}
  \label{fig:siggal}
\end{figure}

Why do galaxies in the high resolution simulations accrete more wind
material? The reason is that at higher resolution, one can resolve
haloes with lower mass in the simulation or haloes at earlier
stages of their assembly when they were still too small to be resolved in
the lower resolution simulations. These low mass haloes launch a
considerable amount of wind material because of their very high mass loading
factors before they grow to sizes that would be resolvable in the
lower resolution simulations. In their subsequent evolution, galaxies in
the high resolution simulations start accreting material from this
earlier generation of winds while the galaxies in the low resolution
simulation are still accreting mostly pristine gas.

The top panel of \fig{fig:siggal} compares the distribution of all
winds that are launched from a simulation as a function of $\siggal$,
which is the velocity dispersion of the haloes from which the winds
are launched. It demonstrates that the high resolution simulation
launches more than twice as much mass in winds as the low resolution
simulation over time and that most of the extra winds are launched from the
lowest mass haloes, i.e., smaller $\siggal$. Furthermore, above the
resolution limit, both simulations launch nearly identical amount of
winds from galaxies that are resolved in both simulations.

The lower panels of \fig{fig:siggal} indicate where the accreted wind
material originates in these two simulations. We select all wind
material that recently accreted unto resolved galaxies (whose halo mass are larger than $10^{10}\msolar$) within the last gigayear and
show the distribution of $\siggal$ for where this wind material originates.
It also demonstrates that the high resolution
simulation not only launches more winds, but also that a considerable
amount of this wind material is recycled.


\section{Summary and Discussion}
\label{sec:discussion}
Modelling the interactions between galactic winds and their surrounding
halo gas has been a challenge for cosmological simulations owing to
the complicated interplay of physical processes that are unresolved in
these simulations. In our previous paper \citep{phewi}, we developed
the Physically Evolved Winds (PhEW) model for analytically evolving cold 
galactic winds through an
ambient medium. Our framework consistently models shocks,
hydrodynamic instabilities, thermal conduction and evaporation, and
makes predictions for the mass loss rate and the deceleration rate of
the cloudlets that match the numerical results of high resolution
cloud-crushing simulations from \citetalias{bs16} under various
conditions.

  In principle, a positive aspect of PhEW is that we can incorporate different models of the small scale physics, e.g., hydrodynamic instabilities, thermal conduction, shattering and entrainment. In this current work, the importance of hydrodynamic instabilities and thermal conduction are encapsulated in the parameters $\fkh$ and $\fS$, respectively. CGM observations could provide important constraints on the parameters, however, detailed comparisons with observations require additional post-processing steps, which we reserve for future work.

In this paper, we provide a detailed numerical prescription for implementing
the PhEW model in cosmological hydrodynamic simulations. The PhEW
model governs the evolution of the wind particles after they were
launched from their host galaxies in a simulation. We performed
several simulations with the PhEW model using various numerical and
physical parameters to study the behaviours of the PhEW particles and
their impact on the stellar and gas properties of the galaxies in the
simulations.

The evolution of PhEW particles in our simulations strongly depends on
halo mass. In low-mass haloes with $\logmvir < 11.5$,
hydrodynamic instabilities dominate mass loss, even though most PhEW
particles are still able to survive for hundreds of Megayears and escape
their host haloes before they disintegrate. PhEW particles survive longest
in intermediate-mass haloes with $\logmvir \approx 12.0$,
because weak thermal conduction can suppress hydrodynamic
instabilities but still does not lead to efficient conductive evaporation.
Some PhEW particles remain in the CGM/IGM even after a
few gigayears. In the massive haloes with $\logmvir > 13.0$, the ambient
halo gas becomes hot enough that strong thermal conduction
leads to quick evaporation of the cloudlets, usually within 200
Megayears. Nevertheless, owing to the high initial speed of these winds, many
particles can still travel a considerable distance, up to 0.5 $\rvir$,
before being fully evaporated.

The behaviour of PhEW particles is designed to be robust to numerical
resolution and hydrodynamic technique (see appendix), in constrast to
many other common galaxy supernova wind
models \citep{katz96, springel03, oppenheimer06, stinson06, dallavecchia12, agertz13, pillepich18a, simba, huang20a},
where wind properties may be more sensitive to resolution and to the complex
interactions between resolution and the hydrodynamic technique.
The PhEW wind behaviour does depend on the PhEW parameters, as it must,
being based on an explicit model of the sub-grid physics.
For example, a smaller cloudlet mass leads
to cloudlets that disintegrate on shorter time-scales in both low-mass and
massive haloes, although choosing a larger value for $\fkh$ helps them to
survive longer in low-mass haloes.

The evolution of wind particles and the galaxy properties in the PhEW
simulations are much more robust to numerics than simulations
without PhEW model. Between the \textsc{gadget-3} and the \textsc{gizmo}
simulations, which rely on different algorithms for solving
hydrodynamics, the behaviours of wind particles are very different even
with the same initial wind speeds and at the same resolution, with the
winds in the \textsc{gizmo} simulation slowing down faster (see appendix).
Furthermore, in both the SPH and the MFM method used by these
simulations, the wind-halo interactions are unresolved and sensitive
to numerical resolution. This leads to very different predictions on
how far the winds can travel, the amount of wind recycling and the
star formation histories of the galaxies. Because of this sensitivity
of the wind model, one often needs to re-adjust the wind parameters
for simulations with different resolutions to obtain similar results
\citep{huang20a}. However, most results from the PhEW simulations
converge very well at different resolutions without the need for wind retuning.
We are hopeful that the PhEW model can be easily adapted to other 
cosmological hydrodynamic simulation codes whether they are Lagrangian, grid based, or moving mesh and still retain the numerical robustness that we demonstrated here.

Our results provide an initial view of how the sub-grid physics modelled by
PhEW affects the evolution of galaxies, winds, and the CGM in simulations
that are matched in numerical resolution, cooling and feedback physics,
and the treatment of wind {\it launch} from galaxies. Our key findings
are as follows.

1. At redshifts $z \geq 1$, the galaxy stellar mass function is little
affected. However, at $z=0$ the mass function is boosted significantly
below $M_* = 10^{11}M_\odot$, achieving a better agreement with observations relative to equivalent simulations without the PhEW \citep[e.g.][]{huang20a}.
This change arises because there is much more wind recycling in these lower mass halos. This increase in recycling is caused by the ability of PhEW particles to shed some of their mass to the ambient CGM before escaping the halo, unlike traditional wind particles.
This recycling leads to significantly better agreement with the
empirically inferred stellar-to-halo mass relation at low redshift and
thus with the observed galaxy stellar mass function. See Figures \ref{fig:gsmfs} and \ref{fig:smhms}.

2. The high-mass end of the galaxy stellar mass function is minimally affected.
In the absence of AGN feedback (or some other physics we have not included),
simulations with or without PhEW produce excessively massive galaxies in
large halos at low redshift, by about a factor of three at the highest masses.
See \fig{fig:gsmfs}.

3. The CGM baryon fraction and the division of CGM baryons between cold and
hot phases is only mildly affected.  Across the halo mass range
$10^{11}-10^{13} M_\odot$ the fraction of baryons in halos is depressed
to 60-80\% of the cosmic baryon fraction in our simulations, at $z=0$
and $z=2$, with or without PhEW. Cold CGM gas exceeds hot CGM gas in
halos below $M_h \approx 10^{11.7} M_\odot$, and hot CGM gas exceeds
cold CGM gas in halos above $10^{12} M_\odot$. Radial profiles of cold
and hot CGM gas are also similar with or without PhEW. See \fig{fig:halomgas}.

4. The distribution of metals in the CGM is radically affected because
PhEW particles share metals with the ambient CGM as their clouds
dissipate, while conventional wind particles retain their metals throughout.
The distribution of CGM gas metallicities with PhEW is skewed but
unimodal, while the conventional wind model gives a very broad and
bimodal distribution. In our PhEW simulations the metallicity of hot
CGM gas and cold CGM gas is similar, while our conventional wind
simulations have a CGM comprised of metal-rich cold gas and metal-poor
hot gas. See Figures \ref{fig:radzprof} and \ref{fig:metalhist}.

Based on these exploratory results, we expect the change from a
conventional wind implementation to PhEW to have an important impact
on predictions for metal-line absorption and emission by the CGM,
and perhaps on X-ray properties of galaxy halos, groups, and clusters.
These changes arise from adopting a physically explicit model for
dispersion of metals from winds to the ambient CGM, rather than a
numerical prescription that is sensitive to resolution and to
implementation details.
We will examine predictions for these informative CGM and galaxy
observables in future work.

\section*{Acknowledgements}
We thank Andrew Benson and Juna Kollmeier for providing computational resources at the Carnegie Institution for Science. We thank the referee, Dylan Nelson, for many useful suggestions. We acknowledge support by NSF grant AST-1517503, NASA ATP grant 80NSSC18K1016, and HST Theory grant HST-AR-14299. DW acknowledges support of NSF grant AST-1909841.

\section*{Data availability}
The data underlying this article will be shared on reasonable request to the corresponding author.

\bibliography{references}

\appendix
\section{A comparison between \textsc{gadget} and \textsc{gizmo} without PhEW}
Simulations that use different numerical algorithms for solving hydrodynamic
equations are known to produce different results in both high resolution
simulations on small-scale hydrodynamic processes \citep{agertz07, hu14, gizmo}
and in large-scale cosmological simulations \citep{frenk99, vogelsberger12, scannapieco12, sijacki12, nelson13, sembolini16b, huang19}.

In this work we implement the PhEW model into \textsc{gizmo} and show how it
affects the properties of galaxies and haloes in \textsc{gizmo} simulations.
The \textsc{gizmo} MFM method solves hydrodynamics very differently from the
\textsc{gadget-3} \citep{springel05} SPH method. \citet{gizmo} shows that the 
two algorithms produce different results in various standard hydrodynamic
test problems. In addition, our version of \textsc{gizmo} also includes 
different cooling and star formation algorithms than we used in our 
\textsc{gadget-3} simulations, as outlined in the main text.
Hence, to isolate changes from our previous published simulations that owe to 
switching from our \textsc{gadget-3} to our \textsc{gizmo}
code, we evolved a non-PhEW simulation using our fiducial \textsc{gadget-3}
model \citep{huang20a} to study how \textsc{gizmo} simulations
compare to \textsc{gadget-3} simulations using the same sub-grid wind model.

The purpose of the appendix is to show how the results from our PhEW simulations might compare to our published work using the \textsc{gadget-3} code. The two codes are different in many aspects so the comparison is not meant to be a convergence study but rather to give an idea to the readers who are familiar with our previously published results how things might have changed with the new code.

This simulation, l50n288-gadget3, represents our
previous simulations that model galactic winds with the sub-grid
model of \citet{huang20a}, and using our previous cooling and star formation
algorithms. Here we use a version of the \textsc{gadget-3} code
\citep{huang19} that employs the pressure-entropy SPH formulation (PE-SPH)
\citep{hopkins13}, the \citet{cd10} artificial viscosity, artificial
conduction \citep{price08}, and a time-step limiter \citep{durier12}.
It models non-equilibrium cooling using the \citet{wiersma09} model
and a \citet{haardt12} UV background and models the interstellar
medium and star formation using the effective equation of
state model of \citet{springel03} based on the observed
\citet{kennicutt98} relation.

In addition, we performed two simulations with \textsc{gizmo} using its implementation of the PE-SPH method instead of the MFM method, one without the PhEW (l50n288-phewoff-sph) and one with the PhEW (l50n288-phew-m5-sph). The PE-SPH method in the \textsc{gizmo} simulations also uses the \citet{cd10} artificial viscosity and artificial conduction as in the \textsc{gadget-3} simulation. We use the same cooling and the star formation algorithms in these simulation as in the other \textsc{gizmo} simulations.

\begin{figure}
  \includegraphics[width=0.85\columnwidth,angle=270]{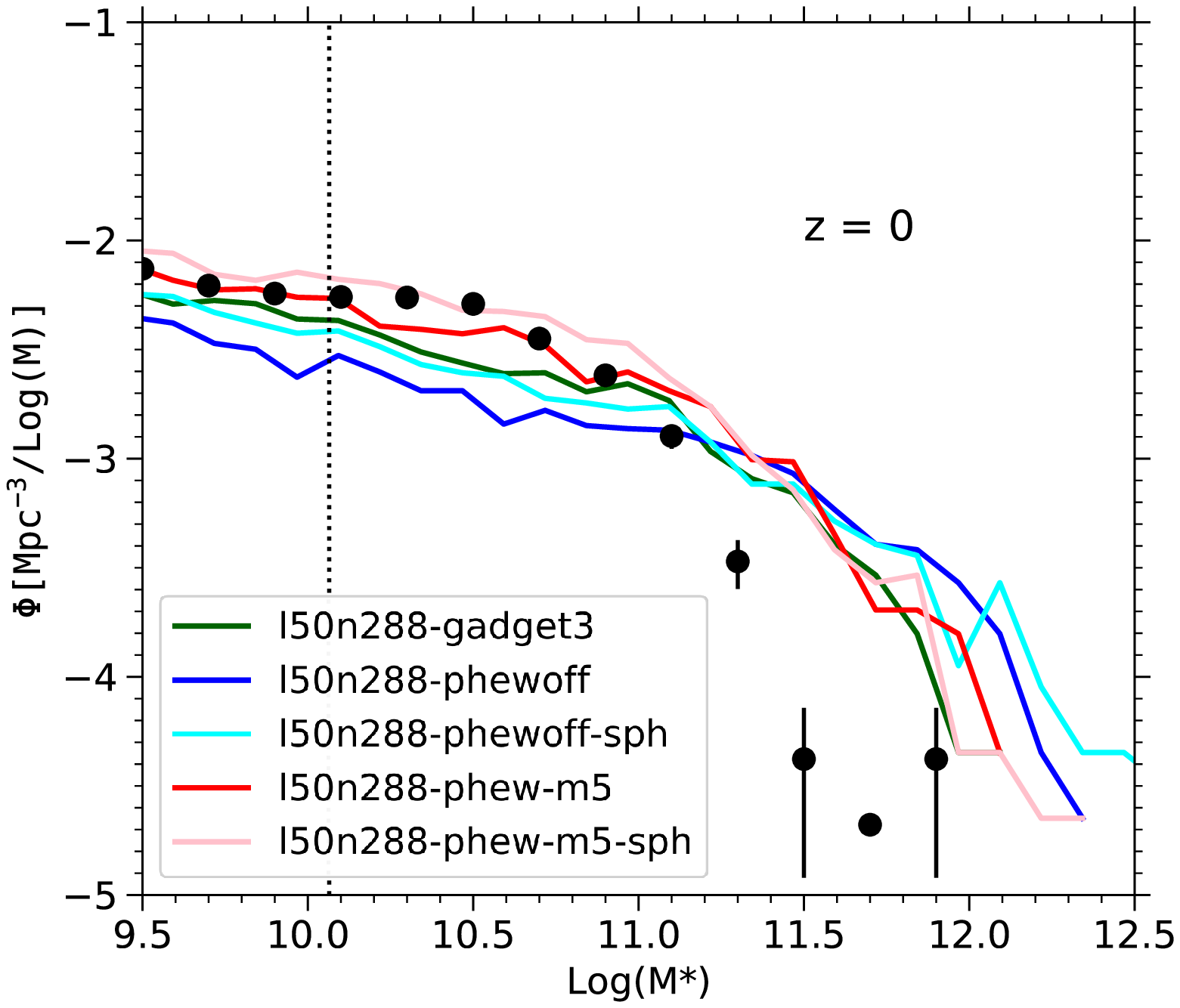}
  \centering
  \caption{The galactic stellar mass functions at $z=0$. Here we compare
the \textsc{gadget-3} simulation (green) with several \textsc{gizmo} simulations, including the l50n288-phewoff (blue, MFM method without the PhEW model), the l50n288-phew-m5 (red, MFM method with the PhEW model), the l50n288-phewoff-sph (pink, PE-SPH method without the PhEW model) and the l50n288-phew-m5-sph simulation (red, PE-SPH method with the PhEW model). The results from the two simulations with the PhEW are close to each other, even though they use different hydrodynamic solvers, while the results from the three simulations without the PhEW are very different.}
  \label{fig:apx_gsmf}
\end{figure}

\begin{figure}
  \includegraphics[width=0.85\columnwidth]{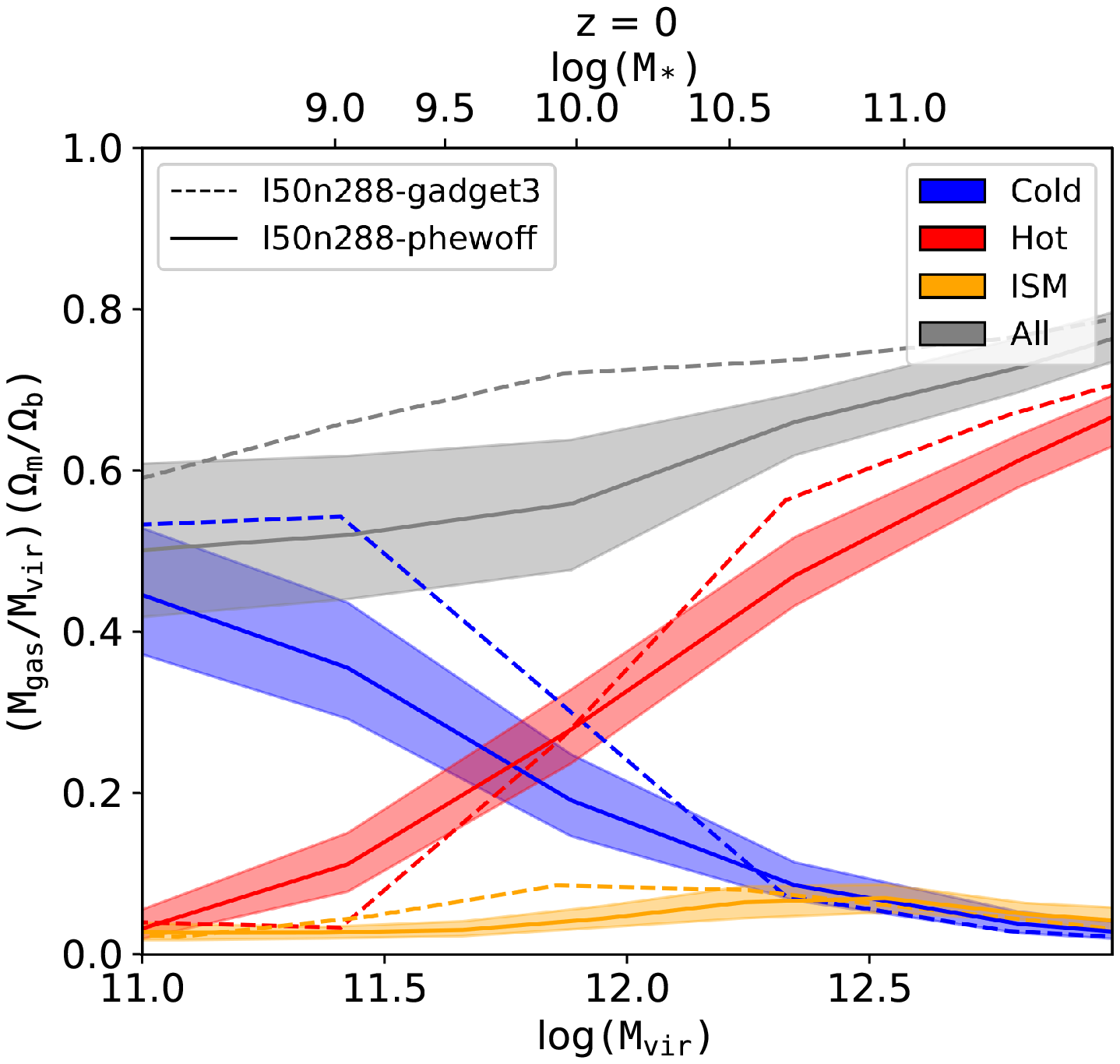}
  \centering
  \caption{The composition of baryons in galactic haloes at $z=0$ similar to
\fig{fig:halomgas}. The dashed and solid lines indicate results from the
\textsc{gadget-3} simulations and the \textsc{gizmo} simulations, respectively.
The shaded area corresponds to the 1$\sigma$ scatter from the \textsc{gizmo}
simulation.}
  \label{fig:apx_halomgas}
\end{figure}

\fig{fig:apx_gsmf} compares the $z=0$ GSMFs from these simulations.
Their GSMFs at higher redshifts ($z \leqslant 1$) are similar to each other. We run all these simulations with the same initial density field but using different numerical methods for solving hydrodynamics, i.e. SPH methods implemented separately in \textsc{gadget-3} and \textsc{gizmo} and the MFM method in \textsc{gizmo}. The GSMFs between the non-PhEW simulations (l50n288-gadget3, l50n288-phewoff and l50n288-phewoff-sph) are different over a wide range of masses. On the other hand, the GSMFs of the PhEW simulations are nearly identical with each other. This shows that when one changes the numerical hydrodynamic method or the numerical resolution in simulations using PhEW it only marginally affects
the $z=0$ GSMF, again demonstrating the numerical robustness of PhEW. However,
when using traditional wind propogation techniques, i.e. not using PhEW, there
are significant differences in the results.

\fig{fig:apx_halomgas} suggests that the cold and hot gas fractions between
the two simulations are similar as well, except that the amount of cold gas
is lower in the \textsc{gizmo} (l50n288-phewoff) simulation.
These results are consistent with
previous findings that the large scale properties of gas and stars from
cosmological simulations are robust to changes in numerical algorithms even if
they may have a large impact on small scale physics \citep{hu14, huang19}.

\begin{figure}
  \includegraphics[width=0.85\columnwidth,angle=270]{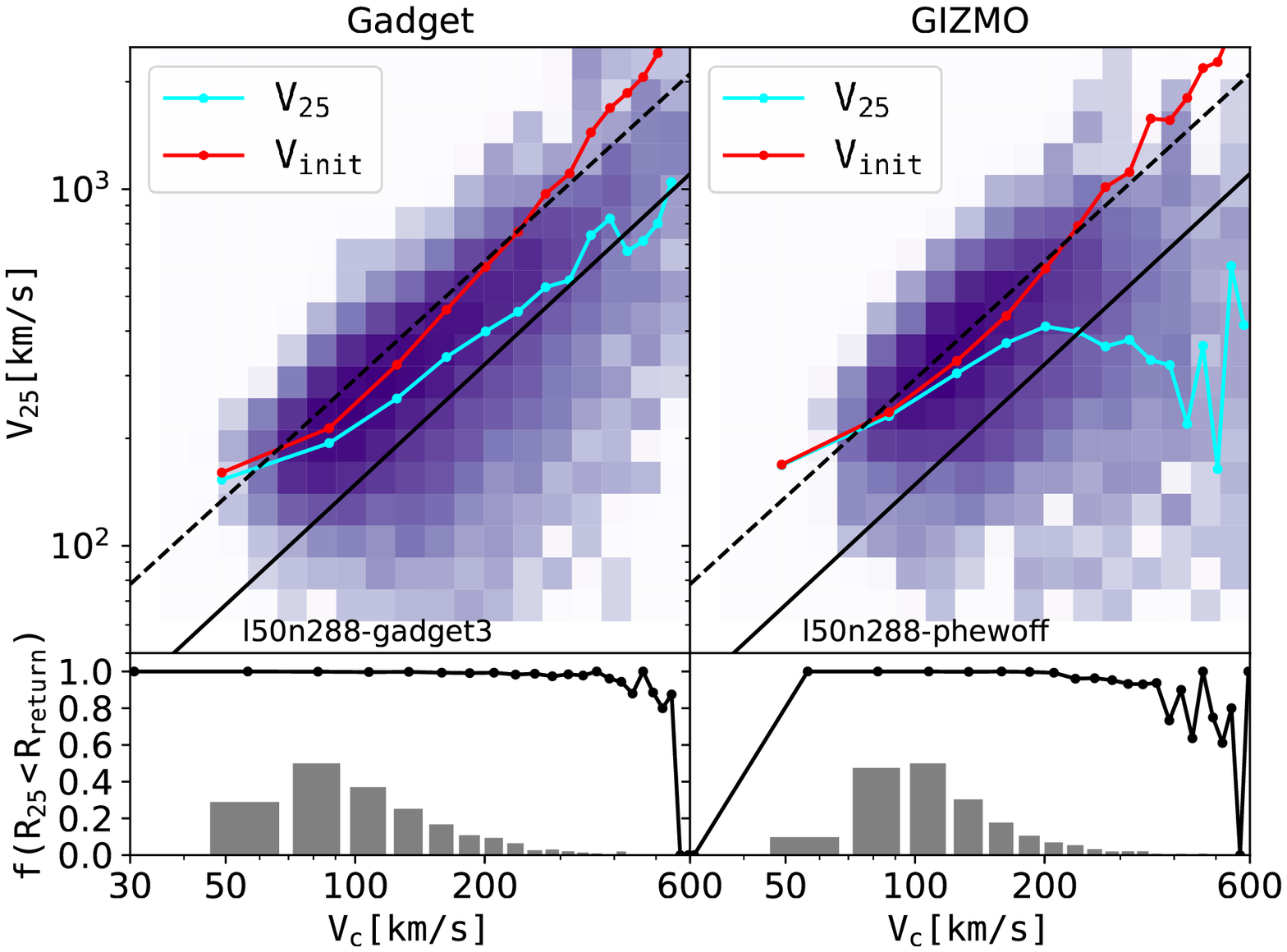}
  \centering
  \caption{The same as \fig{fig:v25vc}, except that we now compare the results
from the \textsc{gadget-3} (l50n288-gadget3)
and the \textsc{gizmo} (l50n288-phewoff) simulations. Both simulations launch winds using the same sub-grid model, but
the different hydrodynamics of the two simulations affect wind propagation very
differently.}
  \label{fig:apx_v25vc}
\end{figure}

However, \fig{fig:apx_v25vc} suggests that our \textsc{gadget-3} versus \textsc{gizmo} simulations have strongly different dynamics, especially in massive haloes. If the different cooling and star formation physics between these two simulations has a minor impact on these dynamics, then the differing treatment of the underlying hydrodynamics may have a strong effect. Even though wind particles in both
simulations have the same initial velocities at any given halo mass, those from
the \textsc{gizmo} (l50n288-phewoff) simulation slow down
faster than in the \textsc{gadget-3} (l50n288-gadget3) simulation. 
As a result, wind particles become
significantly slower in the \textsc{gizmo} (l50n288-phewoff) simulation at 
$0.25\rvir$. Therefore, wind recycling is faster and more frequent in the 
\textsc{gizmo} (l50n288-phewoff) simulation, leading to more
star formation in massive galaxies.

It is important to note, however,  that neither \textsc{gadget-3}
nor \textsc{gizmo} actually resolves the small scale interactions between the
winds and the surrounding gas, making the wind dynamics in both methods suspect
and sensitive to both numerical resolution and numerical artefacts.  
In contrast, the PhEW model is very robust to both the numerical hydrodynamic
algorithm and is almost independent of the numerical resolution.

\bsp
\label{lastpage}
\end{document}